\documentclass[twocolumn]{aastex631}

\usepackage{graphicx}
\usepackage{amsmath,amssymb}
\usepackage{color}
\usepackage{units}
\usepackage{hyperref}
\usepackage{gensymb}
\usepackage{wrapfig,lipsum,booktabs}
\newcommand\um{$\,\rm{\mu m}$}
\newcommand\Lsun{L$_{\rm \odot}$}
\newcommand\Msun{M$_{\rm \odot}$}
\newcommand\Msunyr{M$_{\rm \odot}$\,yr$^{-1}$}

\begin{document} 

\title{Searching Far and Long I: Pilot ALMA 2\,mm Follow-up of Bright Dusty Galaxies as a Redshift Filter} 
\shorttitle{ALMA 2\,mm Follow-up of 850\um{}-bright DSFGs}

\received{15 Nov 2021}
\revised{22 Mar 2022}
\accepted{25 Mar 2022}

\author[0000-0003-3881-1397]{Olivia R. Cooper}
\affiliation{Department of Astronomy, The University of Texas at Austin, 2515 Speedway Boulevard Stop C1400, Austin, TX 78712, USA}

\author[0000-0002-0930-6466]{Caitlin M. Casey}
\affiliation{Department of Astronomy, The University of Texas at Austin, 2515 Speedway Boulevard Stop C1400, Austin, TX 78712, USA}

\author[0000-0002-7051-1100]{Jorge A. Zavala}
\affiliation{Department of Astronomy, The University of Texas at Austin, 2515 Speedway Boulevard Stop C1400, Austin, TX 78712, USA}
\affiliation{National Astronomical Observatory of Japan, 2-21-1 Osawa, Mitaka, Tokyo 181-8588, Japan}

\author[0000-0002-6184-9097]{Jaclyn B. Champagne}
\affiliation{Department of Astronomy, The University of Texas at Austin, 2515 Speedway Boulevard Stop C1400, Austin, TX 78712, USA}

\author[0000-0001-9759-4797]{Elisabete da Cunha}
\affiliation{International Centre for Radio Astronomy Research, University of Western Australia, 35 Stirling Hwy, Crawley, WA 6009, Australia}
\affiliation{ARC Centre of Excellence for All Sky Astrophysics in 3 Dimensions (ASTRO 3D)}

\author[0000-0002-7530-8857]{Arianna S. Long}
\altaffiliation{NHFP Hubble Fellow}
\affiliation{Department of Physics and Astronomy, University of California, Irvine, CA 92697, USA}

\author[0000-0003-3256-5615]{Justin S. Spilker}
\altaffiliation{NHFP Hubble Fellow}
\affiliation{Department of Astronomy, The University of Texas at Austin, 2515 Speedway Boulevard Stop C1400, Austin, TX 78712, USA}
\affiliation{Department of Physics and Astronomy and George P. and Cynthia Woods Mitchell Institute for Fundamental Physics and Astronomy, Texas A\&M University, 4242 TAMU, College Station, TX 77843-4242}

\author[0000-0002-8437-0433]{Johannes Staguhn}
\affiliation{The Henry A. Rowland Department of Physics and Astronomy, Johns Hopkins University, 3400 North Charles Street, Baltimore, MD 21218, USA}
\affiliation{Observational Cosmology Lab, Code 665, NASA Goddard Space Flight Center, Greenbelt, MD 20771, USA}

\begin{abstract}

A complete census of dusty star-forming galaxies (DSFGs) at early epochs is necessary to constrain the obscured contribution to the cosmic star formation rate density (CSFRD), however DSFGs beyond $z \sim 4$ are both rare and hard to identify from photometric data alone due to degeneracies in submillimeter photometry with redshift. Here, we present a pilot study obtaining follow-up Atacama Large Millimeter Array (ALMA) 2\,mm observations of a complete sample of 39 850\um-bright dusty galaxies in the SSA22 field. Empirical modeling suggests 2\,mm imaging of existing samples of DSFGs selected at 850\um{}--1\,mm can quickly and easily isolate the ``needle in a haystack" DSFGs that sit at $z>4$ or beyond. Combining archival submillimeter imaging with our measured ALMA 2\,mm photometry ($1\sigma \sim 0.08$\,mJy\,beam$^{-1}$ rms), we characterize the galaxies' IR SEDs and use them to constrain redshifts. With available redshift constraints fit via the combination of six submillimeter bands, we identify 6/39 high-$z$ candidates each with $>50\%$ likelihood to sit at $z > 4$, and find a positive correlation between redshift and 2\,mm flux density. Specifically, our models suggest the addition of 2\,mm to a moderately constrained IR SED will improve the accuracy of a millimeter-derived redshift from $\Delta z/(1+z) = 0.3$ to $\Delta z/(1+z) = 0.2$. Our IR SED characterizations provide evidence for relatively high emissivity spectral indices ($\langle \beta \rangle = 2.4\pm0.3$) in the sample. We measure that especially bright ($S_{850\rm\mu m}>5.55$\,mJy) DSFGs contribute $\sim10$\% to the cosmic-averaged CSFRD from $2<z<5$, confirming findings from previous work with similar samples.

\end{abstract}

\keywords{Submillimeter astronomy (1735) --- Ultraluminous infrared galaxies (1647) --- Starburst galaxies (1570)}

\section{Introduction}
\label{sec:intro}

A primary goal in extragalactic astronomy is to understand the cosmic star formation rate density (CSFRD) across cosmic time \citep[][]{2014madaudickinson}. The history of cosmic star formation measured at rest frame UV/optical wavelengths currently extends to $z \gtrsim 10$ due to the relative ease of identifying Lyman Break Galaxies (LBGs) in deep \textit{Hubble Space Telescope} and \textit{Spitzer Space Telescope} imaging \citep[e.g.][]{2013ellis,2013oesch,2015bouwens,2015finkelstein,2016finkelstein}. However, measuring solely unobscured star formation by surveying optical and UV emission alone presents an incomplete picture of cosmic star formation, as at least half of the star formation activity in the Universe is obscured by dust \citep{dsfgcaseyreview}.

Dusty star-forming galaxies (DSFGs) dominate cosmic star formation at its peak epoch at $z\sim2-2.5$ \citep*{dsfgcaseyreview}, with extreme star formation rates (SFRs) $\gtrsim 100$ \Msunyr, stellar masses $\gtrsim 10^{10}$ \Msun, and high gas mass fractions $\sim 40-80\%$ \citep*{2013carilliwalter}. While high SFRs imply the presence of young, massive UV-emitting stars, the majority of the UV and optical stellar emission is absorbed and then re-radiated at longer wavelengths in the (sub)millimeter by dust. Therefore, in order to take a full census of cosmic star formation, multi-wavelength observations are necessary to capture both unobscured and obscured star formation. 

Most DSFGs are discovered in wide-area observations using single dish far-infrared/millimeter instruments such as the Submillimeter Common-User Bolometer Array (SCUBA/SCUBA-2) on the James Clerk Maxwell Telescope \citep[JCMT; e.g][]{1997smail,1998hughes,2005chapman,2017koprowski,2019simpson}, the \textit{Herschel Space Observatory} \footnote{\textit{Herschel} is an ESA space observatory with science instruments provided by European-led Principal Investigator consortia and with important participation from NASA.} \citep{2010pilbratt,2010eales,2012oliver}, and the AzTEC instrument \citep{2008scott,2011aretxaga}. Multi-wavelength follow up of hundreds of survey-identified DSFGs reveals that most sit between $1 < z < 3$ \citep[][]{2012casey_zsurvey,2011magnelli,2013magnelli,2013gruppioni,2005lefloch}. While there exists handful of individually studied DSFGs at redshifts as high as $z \sim 5 - 7$ \citep[e.g.][]{2014cooray,2017strandet,2018marrone,2018zavala_z6,2019casey,2020reuter}, such high-$z$ systems have proven difficult to both identify and spectroscopically confirm. This is because (a) DSFGs at $z > 4$ are outnumbered by the dominant DSFG population at $z \approx 1-3$ and (b) there are serious physical and evolutionary degeneracies that make DSFG photometric redshifts highly uncertain \citep[with precision often $\sigma_{\Delta z / 1+z} \gtrsim 1$;][]{mmpz}. This latter point is often seen as a benefit: their strongly negative \textit{k}-correction means that the flux density of DSFGs at $z>1$ remains constant with increasing $z$ for $\lambda_{\rm obs} \gtrsim 850$\um, meaning a DSFG at $z\sim10$ can be observed as readily as a DSFG at $z\sim1$ \citep{2002Blain}. However, when searching for high-$z$ DSFGs, this negative \textit{k}-correction is also a hindrance as it becomes difficult to identify redshifts for galaxies with only long-wavelength emission. This effect is further exacerbated by the (sub)millimeter color degeneracy between dust temperature and redshift. Thus, barring clear identification at other wavelengths, it is easy to confuse $z\sim2$ and $z\sim6$ DSFGs with solely submillimeter observations.

Previous studies lacked the completeness necessary to resolve whether high-$z$ DSFGs play a significant role in cosmic star formation at $z>3 $ \citep[e.g.][]{2018zavala,casey2018b,casey2018a}. For example, \citet{2017koprowski}, \citet{2016rowanrobinson}, \citet[][]{2020dudz}, and \citet[][]{2020gruppioni} all present very different conclusions on the number density of IR luminous galaxies at $z > 3$, due to either limited sample sizes at high redshift or fundamental differences in survey strategy. Additionally, using a blank-field 2\,mm Atacama Large Millimeter/submillimeter Array (ALMA) survey (Mapping Obscuration to Reionization with ALMA, or MORA), \citet[][]{2021zavala} and \citet[][]{2021casey} found that DSFGs contribute $\sim35\%$ at $z = 5$ and only $20-25\%$ at $z = 6 - 7$, suggesting dust-obscured star formation plays a minor role in total star formation at early epochs. On the other hand, \citet[][]{2020gruppioni} find a flatter IRLF slope from $4.5 < z < 6$, with a significant contribution to the CSFRD from obscured sources at high redshift, though again, using a very different survey strategy. While studies like \citet{2021zavala} and \citet[][]{2020gruppioni} provide a good benchmark for the contributions of obscured and unobscured sources to the CSFRD, large uncertainties remain, necessitating a complete census of DSFGs at early epochs.

Empirical modeling by \citet{casey2018b,casey2018a} derived from simulated DSFGs emphasizes the value of longer wavelength observations in easily and cheaply selecting high-$z$ candidates. DSFGs selected at 850/870\um$\,$ with single wavelength measurements can be sorted roughly by redshift using interferometric 2\,mm follow up imaging, which, beyond redshift constraints, simultaneously provides higher precision astrometry due to a $\sim10\times$ smaller beam size \citep[e.g.][]{2021dacunha}. \citet{2014staguhn} report the first deep IRAM/GISMO 2\,mm observations, and detect 15 sources in the Hubble Deep Field with a median redshift of $z = 2.9 \pm 0.9$, higher than the average redshift of 850\um{}-selected DSFGs \citep[e.g.][median $z=2.2$ and an interquartile range $1.7 < z < 2.8$]{2005chapman}. \citet{2019magnelli} present deep GISMO 2\,mm observations in COSMOS and detect four out of five $z > 3$ DSFGs with known (sub)millimeter counterparts in COSMOS catalogs; they suggest 2\,mm surveys favor detection of massive, extremely star-forming, high-$z$ galaxies. Predictions from the SHARK model \citep[][]{2020lagos} are also in line with this observation. The MORA survey --- the largest ALMA 2\,mm blank-field contiguous survey (184 arcmin$^2$) --- demonstrates this, finding an average redshift $\langle z \rangle = 3.6^{+0.4}_{-0.3}$, with 77\% of sources at $z > 3$ and 30\% at $z > 4$, effectively filtering out lower redshift DSFGs \citep[][]{2021casey}. Other millimeter wavelengths have been leveraged to select for higher redshift sources; for example, \citet[][]{2018zavala} conduct a blind search at 3\,mm and detect 16 sources at $>5\sigma$, and \citet[][]{2019williams} serendipitously discover a $z \sim 5-6$ source at similar wavelengths. \citet[][]{2020reuter} find $\langle z \rangle = 3.9$ for a sample of gravitationally-lensed DSFGs selected at 1.4\,mm, where the lensing and millimeter selections combine to filter out low-redshift sources.

In this paper, we present ALMA 2\,mm observations of $>5\sigma$ bright sources ($S_{850\rm\,\mu m} > 5$\,mJy) identified with SCUBA-2. Our goals are to identify the highest-$z$ sources for further follow-up, provide an independent measurement of the volume density of $z > 3$ DSFGs, and test the practical utility of 2\,mm follow up observations as a redshift filter for larger DSFG surveys. We describe the sample and observations in \textsection 2, and in \textsection 3 we present analysis of source redshifts and physical characteristics. \textsection 4 discusses the implications of our measurements, and \textsection 5 summarizes. We assume a Chabrier IMF \citep{2001kroupa} and \textit{Planck} cosmology throughout this paper, adopting $H_0 = 67.7 \rm{\,km\,s^{-1}\,Mpc^{-1}}$ and $\Omega_{\lambda} = 0.6911$ \citep{planck}.

\section{Sample \& Observations}

\subsection{Sample of DSFGs}

In this project, we set out to measure 2\,mm flux densities for a sample of bright 850\um-selected DSFGs. To select our sample, we drew from the 850\um-selected submillimeter galaxies (SMGs) observed with SCUBA-2 as part of the SCUBA-2 Cosmology Legacy Survey \citep[S2CLS;][]{s2cls}. S2CLS, initially conducted starting in 2011 on the James Clark Maxwell Telescope (JCMT), surveyed $\sim5$ square degrees of sky at 850\um. The largest and deepest survey of SMGs at the time, the wide survey component of S2CLS included seven extragalactic fields (UKIDSS-UDS, COSMOS, \textit{Akari}-NEP, Extended Groth Strip, Lockman Hole North, SSA22, and GOODS-North) and reported 2851 submillimeter sources at 850\um$\,$ with a $\geq3.5 \sigma$ detection, with 1313 of those sources at $S_{\rm obs} > 5$\,mJy. Over 90\% of the total survey area reaches a sensitivity of below 2 mJy beam$^{-1}$, with a median depth per field of $\sim 1.2$ mJy beam$^{-1}$ \citep[][]{s2cls}. The S2CLS survey is ongoing; see additional deep imaging in COSMOS presented in \citet{2019simpson}.

Our targets were first selected from the \citet[][]{s2cls} sample of SMGs to lie in ALMA-accessible fields (COSMOS, UDS, and SSA22), and second, to be detected at $S_{850\rm\,\mu m} > 5$ mJy with $>5\sigma$ significance. We selected the brightest 850\um{}-selected DSFGs for 2\,mm follow up due to the observation that brighter DSFGs have a higher average redshift than galaxies at fainter submillimeter flux densities \citep[][]{2015bethermin,casey2018b,casey2018a}. This paper presents a complete sample of 39 such bright DSFGs in SSA22 only; given the depth of SCUBA-2 observations in SSA22, the $>5\sigma$ threshold translates to an effective flux density cut of $S_{850\rm\mu m}>5.55$\,mJy given the field's $\sim1.2\,$mJy rms. The remaining targets in COSMOS and UDS are approved for observations in ALMA Cycle 8 and will be presented in a future work. Though the SSA22 subsample constitutes $\sim 9\%$ of the full sample to be observed, the subsample is sufficiently large to generalize conclusions on the nature of the brightest $\gtrsim5$\,mJy DSFGs.

\subsection{ALMA Observations}

ALMA observations of our sample in SSA22 were taken on 8 January 2020 as one target field of Project \#2019.1.00313.S (PI: Casey). In this field, 39 individual targets were observed with ALMA's Band 4 centered at 2\,mm. ALMA pointings were centered on the reported S2CLS 850\um$\,$ positions \citep{s2cls}. Each target was observed for $<1$ min using the 12-m array, with angular resolution $\sim1\farcs5$ and local oscillator tuning 145\,GHz. The observations had a PWV of 2.05\,mm, with 43 antennae and total on-source time of 34.12\,min ($<1$\,min per source). The phase calibrator was J2217+0220 and the bandpass calibration was J2253+1608. The baseline limit with good phase (80\%) was 279\,m (roughly a C43-2 configuration).

The ALMA images were reduced using the Common Astronomy Software Application (\texttt{CASA}) version 5.6.0.\footnote{\href{https://casa.nrao.edu}{https://casa.nrao.edu}} We adopted robust $=2$ natural weighting to optimize imaging depth given the sources were unresolved at this resolution. The images were primary beam corrected, accounting for the primary beam response decreasing radially outwards from the center of the field. We derive flux density and noise measurements from primary-beam-corrected images using the \texttt{CASA} task \texttt{imstat}. The median rms achieved was $0.08\,$mJy\,beam$^{-1}$ (range  $[0.07 - 0.11]$\,mJy\,beam$^{-1}$), better than the requested rms of 0.1\,mJy. The synthesized beam for these observations is $1\farcs7 \times 1\farcs4$. On these scales ($\sim12 \times 14.5$\,kpc at $z = 2$), it is expected that all detected sources are unresolved, therefore we adopt the peak flux density from \texttt{CASA imstat} within the SCUBA-2 beam (7.5 arcsec radius) as the measured 2\,mm flux density of the source. We confirmed the robustness of this method by comparing to measurements of the source flux using \texttt{CASA} \texttt{viewer} across an extended aperture, and found good agreement between the two measurement methods. 

Given that all of our targets had prior SCUBA-2 detections, we invoked a $3\sigma$ detection threshold for our ALMA observations. While a $3\sigma$ detection threshold is below the nominal $5\sigma$ threshold for blank field detection, other works have demonstrated that prior-based measurements are far less burdened by contaminants \citep[e.g.][]{2013hodge,2017dunlop}. Out of the 39 targets, 35 targets were detected in the 2\,mm observations above this threshold. Only one detected target (SSA.0007) was resolved into two sources, and the remainder were singletons. The 850\um{} and 2\,mm flux densities for all targets are listed in Table \ref{tab:flux}. 

\begin{figure}[h!]
    \centering
    \includegraphics[angle=0,trim=0in 0in 0in 0in, clip, width=0.5\textwidth]{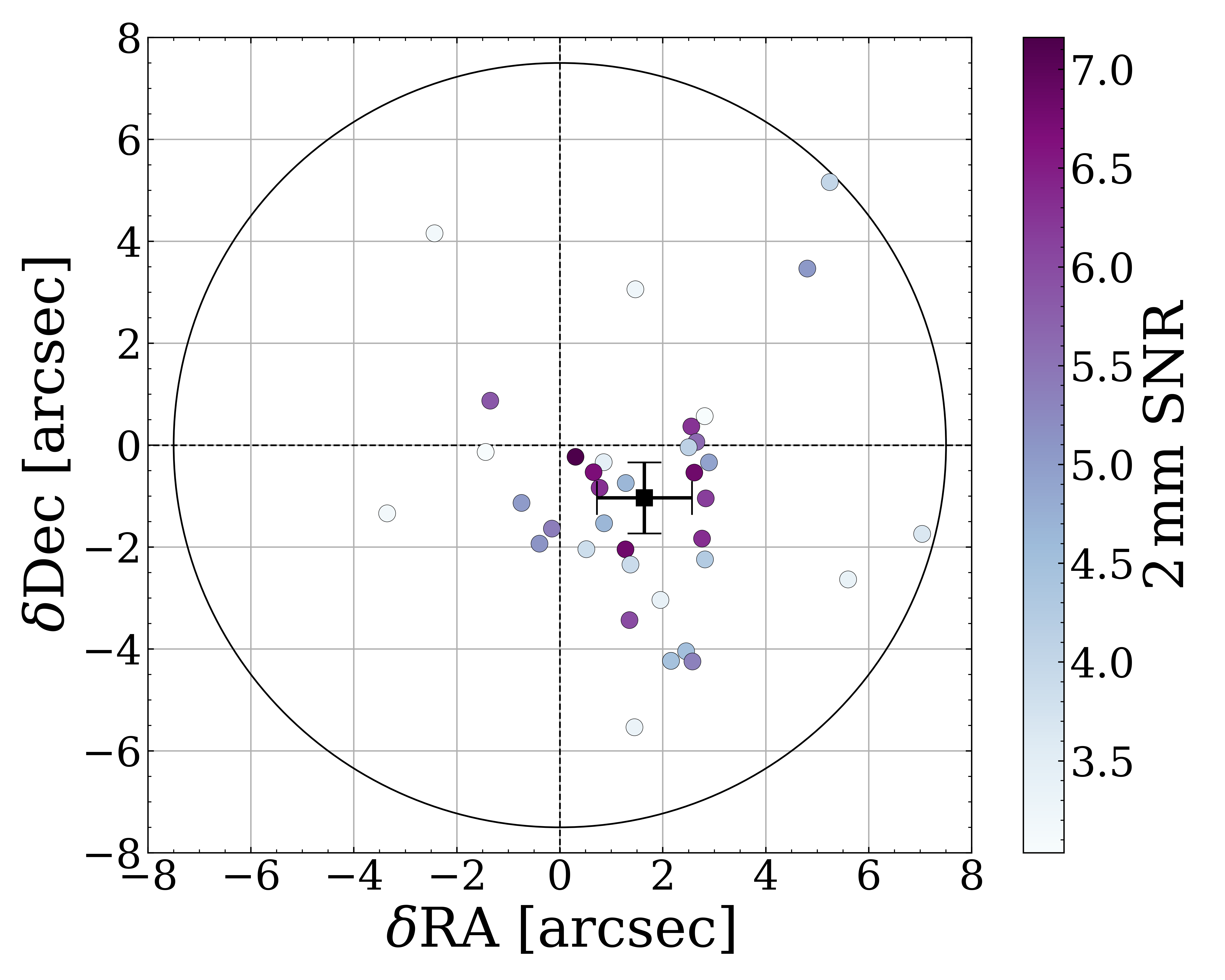}
    \caption{Positional offset between ALMA and SCUBA-2 centroids for the $S_{2\rm{mm}} > 3\sigma$ sources in our sample, colored by single-to-noise ratio (SNR) in ALMA data centered on SCUBA-2 positions (pointing centers). The black square shows the average offset and uncertainties. While there is systematic offset between the SCUBA-2 and ALMA positions, nearly all detected sources lie well within the SCUBA-2 beam FWHM, represented by the black circle. Those far offset tend to have lower SNR suggesting a higher probability of chance alignment.}
    \label{fig:s2asep}
\end{figure}

Comparing S2CLS reported coordinates to our measured coordinates of sources with flux densities $> 3\sigma$, we find an average RA offset of $1\farcs64\pm0\farcs92$ and an average Dec offset of $-1\farcs0\pm 0\farcs7$, for a total net average offset of $1\farcs9\pm 1\farcs2$. Some offset is expected, as SCUBA-2 is a single dish facility and achieves lower resolution and therefore lower astrometric precision than ALMA. We do find some systematic offset as illustrated by Figure \ref{fig:s2asep}, likely originating from the astrometric imprecision of SCUBA-2. Other studies find similar few arcsec systematic offsets between single-dish and ALMA detections \citep[e.g.][]{2013hodge,2015simpson}.

We adopt deboosted flux densities from S2CLS \citep{s2cls} to account for confusion and Eddington boosting, i.e. flux-limited sample bias wherein more sources have fluxes boosted rather than decreased by noise. The median offset between deboosted and observed $S_{850\,\rm{\mu m}}$ flux densities in our catalog is 0.8\,mJy ($\lesssim 20 \%$ of the total flux density), less than the reported uncertainty for each individual source's flux density (which averages to 1.17\,mJy).

\subsection{The SSA22 Field}

The SSA22 field \citep[RA J2000: 22:17:34.6, Dec J2000: +00:16:28.3;][]{1994cowie} was first observed as one of four small selected areas (SSA) chosen for deep multiband imaging followed by complete follow-up spectroscopy. The initial deep imaging was part of the Hawaii K-Band Galaxy Survey, and was first presented in \citet[][]{1994cowie}, with follow-up spectroscopy detailed in \citet[][]{1994songaila} and \citet[][]{1996cowie}. 

The now well-known protocluster structure in SSA22 at $z = 3.09$ was first revealed by \citet{1998steidel}. Protoclusters are rich, overdense regions thought to be the progenitors of massive clusters in the local Universe \citep[][]{2016overzier}. The SSA22 protocluster hosts intense star formation activity, and has an abundance of Lyman break galaxies \citep[LBGs, e.g.][]{1998steidel} and other photometrically-selected galaxies \citep[e.g.][]{2012uchimoto}, Lyman-$\alpha$ emitters \citep[LAEs, e.g.][]{2004hayashino,2012yamada}, Lyman-$\alpha$ blobs \citep[LABs, e.g.][]{2000steidel,2004matsuda,2011matsuda}, and a rare overdensity of DSFGs \citep[e.g.][]{2003stevens,2009tamura,2005geach,2015umehata,2017umehata}. 

Several works have noted that DSFGs may act as signposts for overdense regions or protoclusters at high redshift \citep[\citealt{2016casey}, \citealt{2018lewis}, c.f.,][]{2015miller}. A number of DSFGs have been detected in the SSA22 field, originally discovered with SCUBA-2 observations \citep[e.g.][]{2001chapman,2004chapman_ssa22}, and later detected in AzTEC or ALMA $1.1\,\rm{mm}$ maps \citep[e.g.][]{2009tamura,umehata2014,2015umehata,2018umehata}, some of which have optical to near-infrared and/or millimeter to radio photometric redshifts consistent with the protocluster. \citet{2017umehata,2018umehata} map a small central region (ADF22) of the protocluster with ALMA, revealing an unusually high number of DSFGs at the protocluster core, including intrinsically fainter DSFGs ($S_{1.1\,\rm mm} <1\,$mJy). Of these DSFG protocluster members, very few of those are $S_{850\rm\,\mu m} > 5$\,mJy. Thus, while SSA22 is home to this well known protocluster structure and overabundance of DSFGs, we wish to highlight that the field's population of $S_{850\rm\,\mu m}>5.55\,$mJy DSFGs are not overabundant or more common than the cosmic average based on measurements of bright 850\um{} number counts \citep[e.g.][]{2006coppin,2006scott,2009weiss}. In other words, there is no direct evidence for a statistical excess in the field due to the well-studied overdensity. We explore this further in \textsection4.4.

Note that the presence of an overdensity in the field naturally brings up the thought that gravitational lensing should be considered when measuring bulk characteristics of galaxies behind the protocluster structure. However, the SSA22 protocluster is not yet virialized, and as such its mass distribution is rather diffuse on large scales ($\sim$10\,Mpc), which would lead only to the most modest of weak lensing effects \citep[c.f.][]{2011aretxaga}. In fact, most lines-of-sight across any extragalactic deep field should contain similar extended protoclusters between $2<z<4$ given their volume density implied from measurements of the cluster mass function \citep[e.g.][]{1993bahcall}. Given that the magnitude of lensing caused by this foreground overdensity is likely small and consistent with what is likely present in any extragalactic field, we do not attempt to correct for it here.

\begin{table*}[]
\centering
\begin{tabular}{lcccc}
\hline
\hline
Name      & R.A. & Dec. (J2000)   & $S^{\rm deboost}_{850\,\rm{\mu m}}$ &  $S_{2\rm{mm}}$  \\
       & hms & dms     & mJy beam$^{-1}$ &  mJy beam$^{-1}$  \\
\hline
SSA.0001   & 22:17:32.41 & +00:17:43.8 & 14.5$\pm$1.1  & 0.55$\pm$0.09 \\
SSA.0002   & 22:16:55.62 & +00:28:46.1 & 10.7$\pm$1.4  & 0.26$\pm$0.07 \\
SSA.0003   & 22:16:59.83 & +00:10:39.8 & 10.2$\pm$1.5  & [0.2$\pm$0.07] \\
SSA.0004   & 22:16:51.24 & +00:18:20.7 & 10.0$\pm$1.4    & 0.36$\pm$0.07 \\
SSA.0005   & 22:17:18.79 & +00:18:09.5 & 7.9$\pm$1.3   & [0.23$\pm$0.08] \\
SSA.0006   & 22:18:06.61 & +00:05:20.6 & 8.8$\pm$1.8   & 0.57$\pm$0.08 \\
SSA.0007$\dagger$   & 22:17:37.02 & +00:18:22.6 & 7.2$\pm$1.3   & 0.67$\pm$0.11 \\
SSA.0007.1$\dagger$ & 22:17:37.02 & +00:18:22.6 &     & 0.34$\pm$0.08 \\
SSA.0007.2$\dagger$ & 22:17:36.98 & +00:18:20.61 &    & 0.33$\pm$0.08 \\
SSA.0008   & 22:18:06.46 & +00:11:34.5 & 7.7$\pm$1.5   & 0.21$\pm$0.07 \\
SSA.0009   & 22:17:33.93 & +00:13:52.0 & 7.34$\pm$1.09 & 0.47$\pm$0.07 \\
SSA.0010   & 22:17:01.11 & +00:33:31.4 & 8.9$\pm$2.0     & 0.49$\pm$0.08 \\
SSA.0011   & 22:18:27.88 & +00:25:36.4 & 7.1$\pm$1.5   & 0.53$\pm$0.08 \\
SSA.0012   & 22:17:43.24 & +00:12:31.9 & 7.0$\pm$1.4     & 0.33$\pm$0.07 \\
SSA.0013   & 22:16:57.31 & +00:19:24.0 & 6.9$\pm$1.5   & 0.4$\pm$0.07  \\
SSA.0014   & 22:18:15.26 & +00:19:56.9 & 7.3$\pm$1.4   & 0.59$\pm$0.09 \\
SSA.0015   & 22:16:52.19 & +00:13:42.3 & 6.8$\pm$1.5   & 0.44$\pm$0.08 \\
SSA.0016   & 22:18:06.19 & +00:04:01.5 & 8.7$\pm$2.0     & 0.44$\pm$0.07 \\
SSA.0017   & 22:17:44.04 & +00:08:21.8 & 6.5$\pm$1.5   & 0.21$\pm$0.07 \\
SSA.0018   & 22:17:28.31 & +00:20:26.2 & 6.3$\pm$1.3  & 0.52$\pm$0.08 \\
SSA.0019   & 22:17:42.26 & +00:17:02.3 & 6.0$\pm$1.4     & 0.36$\pm$0.07 \\
SSA.0020   & 22:18:27.20 & +00:19:31.3 & 5.6$\pm$1.5   & 0.25$\pm$0.08 \\
SSA.0021   & 22:17:41.35 & +00:26:41.5 & 5.8$\pm$1.4   & 0.36$\pm$0.07 \\
SSA.0022   & 22:18:23.61 & +00:26:33.1 & 6.0$\pm$1.4     & 0.38$\pm$0.08 \\
SSA.0023   & 22:18:17.21 & +00:29:32.7 & 5.8$\pm$1.3   & 0.31$\pm$0.08 \\
SSA.0024   & 22:18:10.12 & +00:15:55.2 & 5.7$\pm$1.4   & [0.18$\pm$0.07] \\
SSA.0025   & 22:17:09.51 & +00:14:08.9 & 5.5$\pm$1.5   & 0.38$\pm$0.08 \\
SSA.0026   & 22:18:13.51 & +00:20:31.2 & 5.5$\pm$1.3   & 0.59$\pm$0.09 \\
SSA.0027   & 22:16:32.20 & +00:17:46.4 & 5.6$\pm$1.4   & 0.39$\pm$0.09 \\
SSA.0028   & 22:16:50.06 & +00:22:48.4 & 5.3$\pm$1.3   & 0.35$\pm$0.08 \\
SSA.0029   & 22:18:27.10 & +00:21:36.4 & 5.13$\pm$1.25 & 0.28$\pm$0.07 \\
SSA.0030   & 22:18:33.06 & +00:18:42.2 & 5.41$\pm$1.23 & 0.24$\pm$0.08 \\
SSA.0031   & 22:17:17.43 & +00:31:37.4 & 5.1$\pm$1.4   & 0.26$\pm$0.08 \\
SSA.0032   & 22:17:32.31 & +00:29:30.7 & 5.1$\pm$1.3   & 0.27$\pm$0.08 \\
SSA.0033   & 22:17:03.52 & +00:26:03.7 & 4.9$\pm$1.3   & 0.25$\pm$0.08 \\
SSA.0034   & 22:17:02.27 & +00:15:53.9 & 4.9$\pm$1.3   & 0.23$\pm$0.07 \\
SSA.0035   & 22:18:06.72 & +00:06:30.9 & 5.6$\pm$1.4   & 0.65$\pm$0.10 \\
SSA.0036   & 22:17:31.78 & +00:14:54.5 &
4.9$\pm$1.3   & [0.20$\pm$0.07] \\
SSA.0037   & 22:18:29.06 & +00:08:35.0 & 6.3$\pm$1.9   & 0.25$\pm$0.07 \\
SSA.0038   & 22:17:02.95 & +00:24:39.2 & 4.8$\pm$1.3   & 0.29$\pm$0.07 \\
SSA.0041   & 22:18:34.99 & +00:21:42.6 & 6.0$\pm$1.3     & 0.45$\pm$0.08 \\ 

\hline
\hline
\end{tabular}
\caption{All SSA22 targets observed in Band 4 with ALMA. Positions are in J2000, with 2\,mm detected source positions from \texttt{CASA imstat} measurements. $S^{\rm{deboost}}_{850\,\rm{\mu m}}$ is the 850\um\ flux density in mJy beam$^{-1}$ reported from S2CLS \citep[][]{s2cls} and $S_{2\rm{mm}}$ is our measured 2\,mm flux density in mJy beam$^{-1}$. All sources with S/N $> 3$ from ALMA are considered detections given the SCUBA-2 detection as a prior; flux densities of non-detections at 2\,mm are listed in brackets. $\dagger$SSA.0007 is resolved into two components with ALMA (SSA.0007.1 and SSA.0007.2), but is unresolved with SCUBA-2, so $S_{2\rm{mm}}$ for SSA.0007 is reported as the sum of the two components. }
\label{tab:flux}
\end{table*}

\subsection{Redshifts from the Literature}

We crossmatch spectroscopic and photometric redshifts from literature on the SSA22 field to our sample. Positions used for crossmatching are derived from 2\,mm measurements for sources detected at $>3\sigma$ in 2\,mm, otherwise S2CLS reported positions are adopted. The target SSA.0001 is a confirmed protocluster member reported in \citet{umehata2014} with a spectroscopic redshift of $z = 3.092$. SSA.0007 is resolved into two components in our 2\,mm data and is also spectroscopically confirmed as part of a dense group within the protocluster at $z = 3.09$, possibly in a multiple merger phase \citep{2015umehata,2016kubo,2017umehata,2018umehata,2019umehata}. Though it is resolved into two components with ALMA, in our SED characterization we treat SSA.0007 as one source since it is unresolved in other bands. Therefore, we adopt the sum of the 2\,mm flux densities for SSA.0007.1 and SSA.0007.2 reported in Table \ref{tab:flux} for our SED characterization described in Section \textsection3. As SSA.0007.1 and SSA.0007.2 are at the same redshift and thus physically associated, taking the sum of their photometry and treating SSA.0007 as one source corresponds to the net emission from the physically-associated pair and avoids contamination from sources at other redshifts. In addition to the sources confirmed in the protocluster, three galaxies are confirmed spectroscopically to be in the foreground of the protocluster: SSA.0009 at $z = 2.555$ \citep{2005chapman}, SSA.0019 at $z = 2.278$ \citep{2012alaghbandzadeh}, and SSA.0031 at $z = 2.6814$ (confirmed via Lyman-$\alpha$ emission; Cooper et al. in prep). About a third of our targets (14/39) also have optical-infrared (OIR) photometric redshifts and about one quarter (9/39) have mm/radio photometric redshifts (beyond those we calculate herein), both reported in \citet{umehata2014}. In \textsection 3.1, we analyze protocluster membership for our sample relative to likely foreground and background sources.

\subsection{Ancillary Data}

As noted in Section \textsection2.4, for sources detected at $>3\sigma$ in 2\,mm, positions derived from 2\,mm measurements are adopted, otherwise we take the S2CLS positions. The search radius was informed by the worst resolution being crossmatched -- typically the resolution for the ancillary data or the S2CLS resolution of 7.5 arcsec if the source's S2CLS position was used. For each ancillary photometric dataset, counterpart identification was verified at the corresponding wavelength to ensure the counterpart was not an image artifact or noise. 

We crossmatch our sample to the X-ray catalog for SSA22 presented in \citet{2009lehmer} to check for the presence of luminous active galactic nuclei (AGN). AGN could be a concern in using millimeter photometry to constrain redshifts, as AGN can heat ISM dust to temperatures warmer than galaxies without AGN. Fifteen sources have X-ray coverage but just two sources (SSA.0001 and SSA.0007) have significant X-ray luminosities (0.5–8\,keV luminosities within $10^{43} - 10^{45}\,\rm erg\,s^{-1}$) indicating potentially powerful AGN activity for those targets. This suggests the sample as a whole is not dominated by X-ray bright AGN (extremely obscured AGN could produce hotter dust temperatures and cannot be ruled out without follow-up data from e.g. \textit{JWST}), and that the presence of AGN is unlikely to impact the millimeter photometric analysis presented later in this paper as both SSA.0001 and SSA.0007 have secure spectroscopic redshifts. 

We crossmatch our SSA22 sources with reported AzTEC 1.1\,mm flux densities from \citet{umehata2014} and find $\sim70\%$ (27/39) of our sample is detected down to SNR of 3.8. This threshold corresponds to a flux density detection limit of $S_{1.1\,\rm mm} \gtrsim 2.45 - 4.55\,\rm mJy\,beam^{-1}$. Flux densities at 1.1\,mm for the other 12 sources were extracted from the AzTEC map at the best available positions (precise ALMA 2\,mm positions for 2\,mm detections, else SCUBA-2 850\um{} positions), with one source (SSA.0010) lying outside the AzTEC field. 

\textit{Herschel}/SPIRE measurements of SSA22 at 250\um{}, 350\um{}, and  500\um{} were extracted using cross-matched positional priors from MIPS 24\um{} data \citep{swinbank2014,2016kato}. Note that we do not use the MIPS 24\um{} data directly in our analysis because the depth varies significantly across the field and the coverage is incomplete. Furthermore, at the expected redshifts of our sources ($z \approx 2 - 6$), observed 24\um{} does not probe rest-frame FIR emission, but instead measures the mid-IR, rich with complex Polycyclic Aromatic Hydrocarbon (PAH) emission and silicon absorption that we are not attempting to constrain in this work. SPIRE maps were deblended by oversampling to $1$"/pixel from the original \textit{Herschel}/SPIRE $8$" pixel maps, and the extended cirrus emission from the Milky Way's interstellar medium was subtracted. About half of our sample (20/39) overlaps with the MIPS 24\um{} coverage, and from this subset we identified 18/39 counterparts within the MIPS 24\um{} FWHM of 7". For another 19 sources, we directly extract flux densities from deblended \textit{Herschel}/SPIRE maps using our best positional constraints. The flux densities of the final two sources were extracted from the point source \textit{Herschel}/SPIRE maps (optimized for point source extraction, ideal for our unresolved sources), as they lay outside the \citet{swinbank2014} catalog and maps. We compared flux densities of overlapping sources from the point source \textit{Herschel}/SPIRE maps and the deblended maps/catalog, and find no statistical differences within uncertainty. All of our measurements from the \textit{Herschel} data are confusion limited \citep{spire}.

For sources without positional priors in the MIPS 24\um$\,$ catalogs (i.e. those with flux densities extracted from the deblended maps directly) we conduct a statistical analysis of cataloged sources from \citet{swinbank2014} to infer an appropriate rms for our \textit{Herschel} flux density measurements. For each flux density measurement, we calculate the 68\% confidence interval of the catalog rms values within a 0.01 dex bin centered on the measured flux density. As the data are confusion limited, we adopt the confusion error from \citet{spire} for any sources with derived errors less than these confusion error limits. This is a conservative estimate of the noise given the uncertainty in positional priors of deblended catalogs; in other words, at or below the confusion limit, confidence in the accuracy of positional priors is greatly reduced.

For our photometric redshift and SED fitting, we add absolute flux scale calibration uncertainty in quadrature with the statistical photometric uncertainties, taking that calibration error to be 4\% for \textit{Herschel} \citep[][]{2013bendo}, 5\% for SCUBA-2 \citep[][]{2020simpson}, 10\% for AzTEC \citep[][]{2008wilson}, and 5\% for ALMA \citep[][]{2021braatz}.

\section{Redshift and FIR SED Analysis}

In this section, we combine existing FIR/mm photometry with new 2\,mm data to identify the highest redshift candidates and constrain dust SEDs for our targets.

\subsection{Photometric Redshifts}

We derive a millimeter-based photometric redshift using the \texttt{MMpz} tool for each galaxy in our sample, as detailed in \citet{mmpz}. \texttt{MMpz} uses rest-frame FIR/mm reprocessed dust emission to derive a photo-$z$ probability density distribution. The probability distributions are based on the measured distribution of galaxy SEDs in the empirical relation between rest-frame peak wavelength and total IR luminosity, i.e. the $L_{\rm IR}$-$\lambda_{\rm peak}$ plane. The technique accounts for instrinsic SED breadth as it probes a wide range of dust temperatures at fixed IR luminosity. Estimating redshifts in the long wavelength regime suffers from a strong degeneracy between (sub)millimeter flux density and colors
with redshift; this algorithm is suited for data like these that lack other redshift constraints from spectroscopy or OIR photometry. Further, here we target a bright subset of DSFGs, a regime where OIR photo-$z$ estimates can often be less accurate given differential attenuation with wavelength due to complex geometries \citep[][]{2012casey_z,2015dacunha}. Note that \texttt{MMpz} assumes the dust emissivity spectral index $\beta = 1.8$, which is a parameter fit during the dust SED characterization described in \textsection 3.2. Fixing $\beta$ in this way has a minimal impact for use in photometric redshift fitting given the large uncertainties on redshift constraints from temperature degeneracies; see \citet[][]{mmpz} for further discussion.

Comparing \texttt{MMpz} photo-$z$ results to 12 OIR photo-$z$'s from \citet{umehata2014}, we calculate $\Delta z/ (1+z) = 0.13$ with no systematic offset. Note that our comparison to OIR photo-$z$'s excludes SSA.0028, which has a reported OIR photo-$z$ of $z = 0$, inconsistent with both our FIR results and publicly available shallow SDSS optical imaging, which shows no obvious local universe source within 12'' \citep[][]{sdss_dr16}. Additionally, we exclude source SSA.0006 which has an OIR photo-$z$ of $z = 6$; its detection at 24\um{} likely precludes such a high redshift solution. While the sample of galaxies with existing spectroscopic redshifts is much smaller (limited to SSA.0001, SSA.0007, SSA.0009, SSA.0019, and SSA.0031), we find good agreement of $\Delta z/ (1+z) = 0.04$ between our \texttt{MMpz} results and spectroscopic redshifts.

In order to characterize bulk properties of these galaxies in the absence of precise redshift information, we use \texttt{MMpz} photo-$z$ probability density distributions to sort the sample into three redshift bins, with the middle bin at $2.6 < z < 3.6$, centered on the SSA22 protocluster redshift of $z = 3.1$. Note that while we position this mid-$z$ bin at the protocluster redshift, we set a bin size much larger than the spectroscopically confirmed protocluster redshift range as we do not expect the presence of the protocluster to bias our sample; we quantify this further in Section \textsection4.4. To bin the sample, we integrate the PDF in intervals of width $\Delta z = 0.1$ and then consider the 3 highest probability redshift intervals. If any of the redshift intervals with highest probability are within $\Delta z = 0.5$ of $z = 3.1$, the galaxy is categorized in the mid-$z$ bin. If the highest probability redshift intervals are lower or higher than the mid-$z$ bin, they are binned as low-$z$ or high-$z$, respectively. In a couple of cases, the 3 highest probability redshift intervals span multiple bins, with only one in the mid-$z$ bin (e.g. mid-$z$ and low-$z$, or mid-$z$ and high-$z$). In these cases we categorize the galaxy in the mid-$z$ bin. This binning technique accounts for the width of the photo-$z$ PDF, rather than only considering the PDF peak. This results in low-$z$-binned sources ranging $1.2 < z < 2.6$, mid-$z$-binned sources ranging $2.5 < z < 3.6$, and high-$z$-binned sources ranging $3.4 < z < 4.7$.

\begin{figure}[h!]
    \centering
    \includegraphics[angle=0,trim=0in 0in 0in 0in, clip, width=0.5\textwidth]{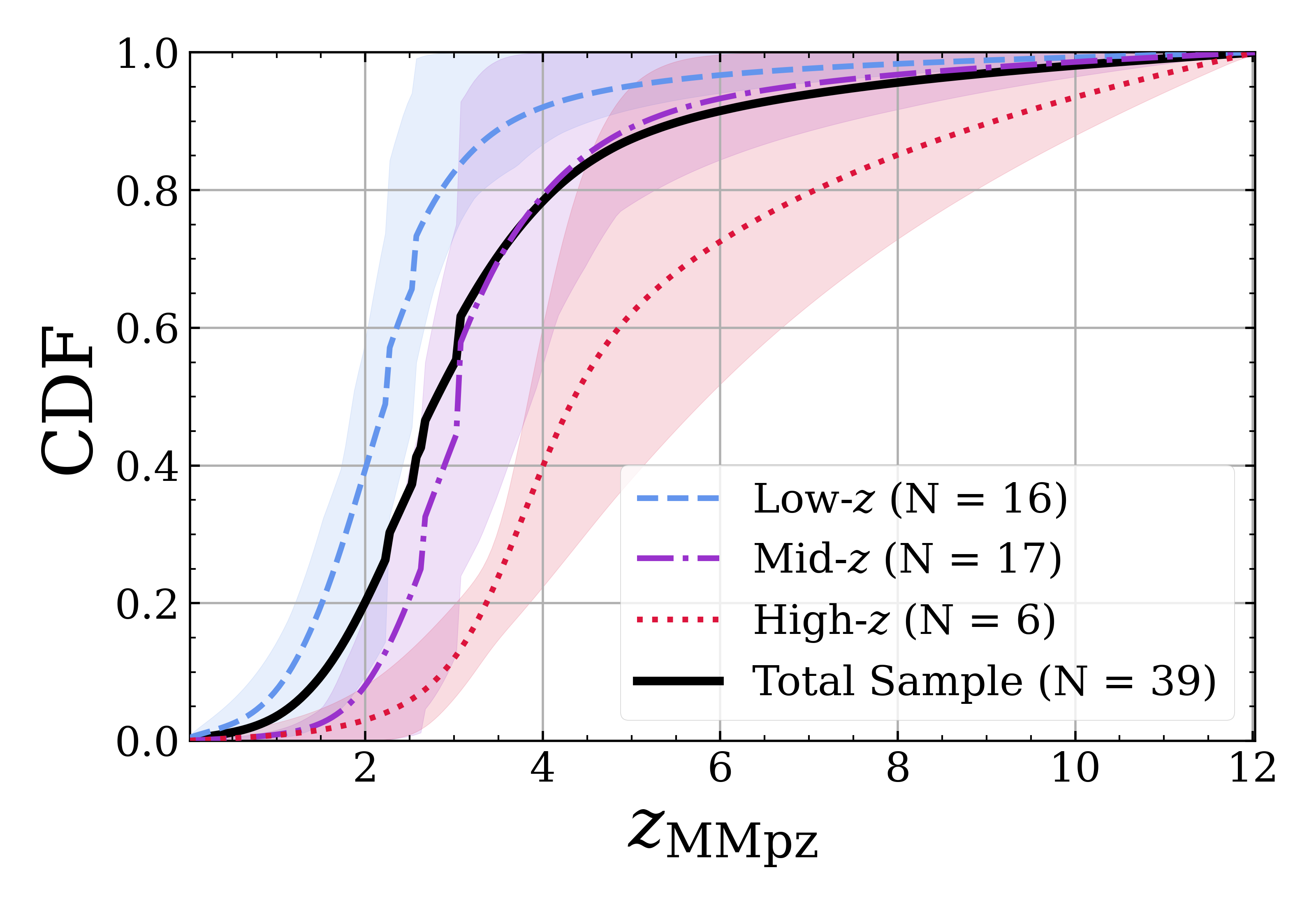}
    \caption{Redshift cumulative probability distribution functions (CDFs) from \texttt{MMpz}, with any spectroscopic redshift represented as delta functions at the measured redshift. The full sample is shown in black, and in blue, purple, and red are the CDFs for the low redshift, mid redshift, and high redshift bins, respectively. The shaded regions show the 68\% confidence interval for each redshift bin based on the individual \texttt{MMpz} fits. For the high-$z$ sample, $\sim 60\%$ of the aggregate probability distribution is at $z > 4$.}
    \label{fig:zbins}
\end{figure}

From our millimeter photo-$z$ measurements and ancillary redshift information, we find 16 low-$z$ galaxies with median $\langle z_{\rm{low}-z} \rangle = 2.08 \pm 0.16$, 17 mid-$z$ galaxies with median $\langle z_{\rm{mid}-z} \rangle = 2.98 \pm 0.11$, and 6 high-$z$ galaxies with median $\langle z_{\rm{high}-z} \rangle = 3.8 \pm 0.3$, as shown in Figure \ref{fig:zbins}. For the full sample, we find a median $\langle z \rangle = 2.7 \pm 0.2$. Errors on the median redshifts are derived from bootstrapping. Similarities between the derived median redshift for this sample and for similar 850\um{}-selected samples of galaxies in the literature are discussed in Section \textsection 4.1.

Note that \texttt{MMpz} takes both color and luminosity into account, and presumes no redshift evolution of $\lambda_{\rm peak}$ (and therefore, $\rm T_{\rm dust}$). Therefore, we expect $z_{\rm MMpz}$ to have some correlation with 2\,mm flux density. This correlation, as well as the utility of 2\,mm flux densities as a redshift filter, is discussed further in \textsection4.2.

\subsection{IR SED Characterization}

Most galaxies in the sample lack precise redshift constraints; for sources lacking spectroscopic redshifts, we use photometry to derive a redshift from \texttt{MMpz}. We then fix the best available redshift solution to infer basic characteristics about the galaxies' SEDs. Note this degeneracy limits the scope of our conclusions as it may lead to underestimated uncertainties. Since the new 2\,mm data is on the Rayleigh-Jeans tail, we focus our analysis on that portion of the spectrum. 

Each galaxy's FIR/mm SED is fit to a modified blackbody added piecewise with a mid-infrared power law. We utilize a technique similar to that described in \citet{2012casey}, but replace least-squares fitting with Bayesian analysis; the full tool will be presented in a forthcoming publication (Drew \& Casey, submitted). The mid-infrared power law is joined to the modified blackbody at the point where the blackbody slope is equal to the power law index $\alpha_{\rm MIR} = 2$ \citep[consistent with other works, e.g.][]{2010kovacs,2012casey,2012u}. The general opacity model is assumed, where the optical depth ($\tau$) equals unity at $\lambda_{\rm rest} = 200$\um{} \citep[e.g.][]{2011conley,2012greve}. Best fit SEDs are found based on a Markov Chain Monte Carlo convergence. 

Input into the IR SED fitting routine is the best available redshift prior (where an existing spec-$z$ is preferred, else an OIR photo-$z$ if available, otherwise fixed to the $z_{\rm MMpz}$ solution) and the FIR/mm photometry (at 250\um{}, 350\um{}, 500\um{}, 850\um{}, 1100\um{}, and 2000\um{}) with associated uncertainties for the flux density measurements. Our fixed parameters are $\alpha_{\rm MIR} = 2$ and $\lambda_0 = 200$\um{} (the wavelength where optical depth equals unity) near the intrinsic peak of the dust SED. As neither can be directly constrained in this dataset, these broad population averaged values are assumed. We do include a CMB correction term in our fitting procedure to account for ISM dust heating from the CMB at high redshift \citep[][]{2013dacunha}, however, since the galaxies in our sample are predominantly at $z < 5$, this CMB correction is quite small for our sample in practice.

We used the SED fitting algorithm to find the best fit SED with measurements for each of the following free parameters: emissivity spectral index ($\beta$), total infrared luminosity ($L_{\rm IR}$, taken from 8-1000\um), dust temperature ($T_{\rm dust}$), and rest-frame peak wavelength ($\lambda_{\rm peak}$). The last two variables have a fixed relationship given our assumed opacity model with $\lambda_0 = 200$\,\um{} (see Figure 20 of \citealt*[][]{dsfgcaseyreview}).

\begin{figure}[]
    \centering
    \includegraphics[angle=0,trim=0in 0in 0in 0in, clip, width=0.5\textwidth]{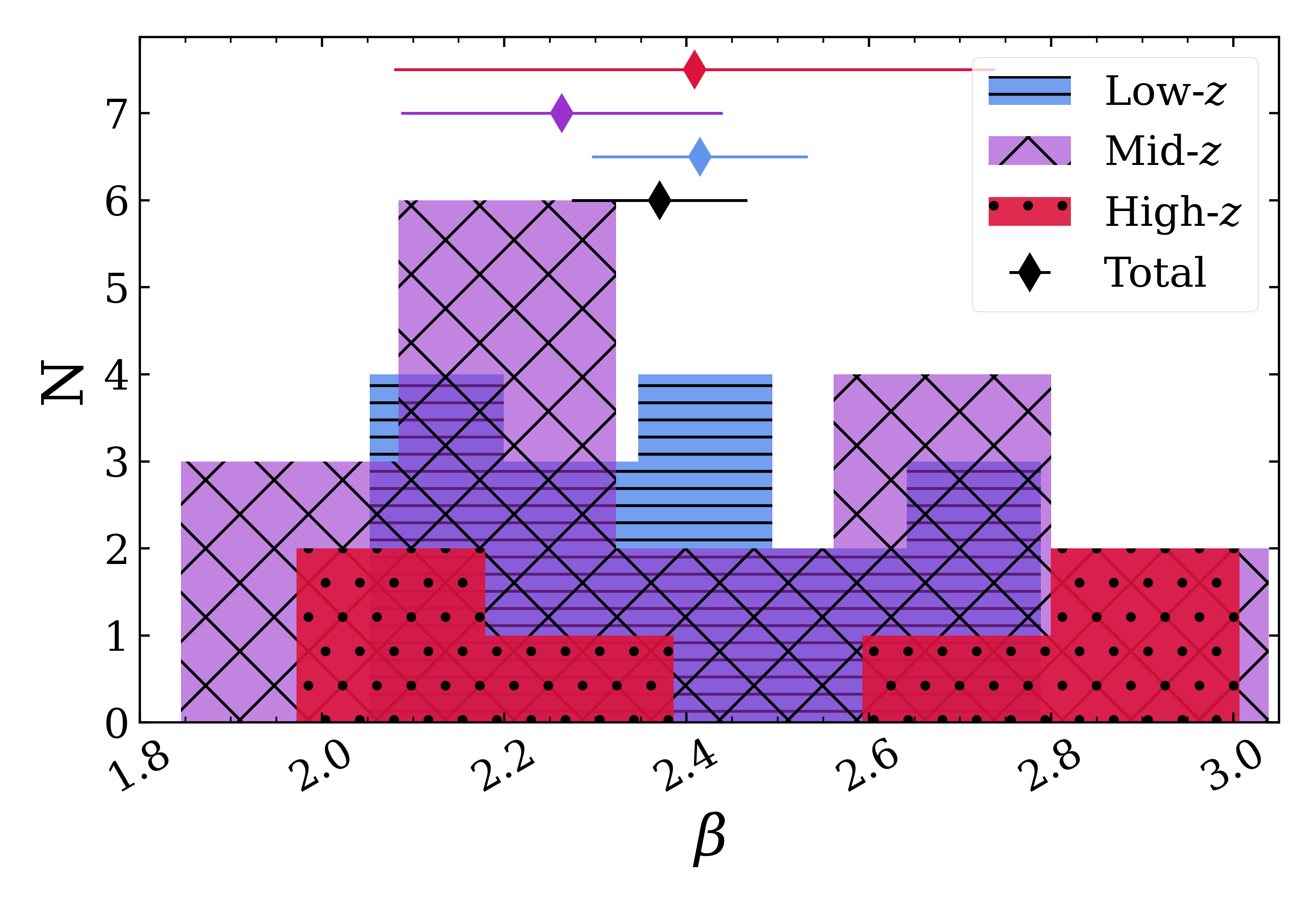}
    \caption{A histogram of measured emissivity spectral index ($\beta$) grouped by redshift bin. Low redshift sources are shown in the blue striped region, mid redshift sources in the purple hased region, and high redshift sources in the red dotted region. The median $\beta$ with uncertainty is plotted as diamonds colored by redshift bin above the histograms, with the total median shown as a black diamond. All three distributions are consistent within error with median $\beta = 2.4 \pm 0.3$.}
    \label{fig:beta}
\end{figure}

Overall, our median sample properties are comparable to other samples of DSFGs. As our targets are selected for being among the brightest 850\um-selected DSFGs in the SCUBA-2 survey, we find generally high IR luminosities such that all may be categorized as ultraluminous IR galaxies (ULIRGs, $L_{\rm IR} \geq 10^{12}$\,\Lsun), with a median of $\rm log\,L_{\rm IR} = 12.7 \pm 0.2$. Our sample is comparable to other samples with IR luminosities a few times $10^{12}$\,\Lsun \citep[][]{2020dudz,2015dacunha,2017miettinen}. 

We convert our $L_{\rm IR}$ measurements to a SFR using the TIR calibrator from \citet{2011hao} and \citet{2011murphy}, consistent with the review \citet{2012kennicuttevans}. We find characteristically high SFRs $\sim 500 - 1000$ \Msunyr, with median SFR = $700 \pm 400$ \Msunyr. While this is somewhat higher than other studies (e.g. 290\,\Msunyr in \citealt{2020dudz}, and 280\,\Msunyr in \citealt{2015dacunha}), we expect higher SFRs corresponding to our bright $S_{850\rm\,\mu m} > 5.55\,$mJy sample selection (the effective flux density cut for our criteria of $S_{850\rm\mu m}>5$\,mJy detection at $>5\sigma$ in SSA22 given the $\sim1.2\,$mJy rms). For comparison, \citet[][]{2020dudz} make a flux density cut $S_{850\rm\mu m}\geq3.6$\,mJy and \citet{2015dacunha} make a flux density cut of $S_{870\rm\mu m}\geq3.5$\,mJy. We discuss the cosmic-averaged contribution to the CSFRD for the sample in Section \textsection4.4.

Peak wavelength measurements span 70\,\um{}$ < \lambda_{\rm peak} < 149$\,\um{}, and consequently dust temperatures span $27 \rm\, K < \rm T_{\rm dust} < 72$ K, with a median $\rm T_{\rm dust} = 45 \pm 8$\,K. \citet{2020dudz} find a slightly lower median dust temperature of 30.4\,K for their sample, with no evolution in $\rm T_{\rm dust}$ at constant IR luminosity from $1.5 < z < 4$. \citet{2021dacunha} also find a median $T_{\rm dust} = 30^{+14}_{-8}$ K. We note that $T_{\rm dust}$ is not consistently defined and relies on assumptions of $\beta$ and $\lambda_0$, which are degenerate with $T_{\rm dust}$ \citep[e.g. Figure 6 in][]{2016spilker}. Further, \citet[][]{2021dacunha} find that optically-thin models tend to bias the temperature low compared with optically thick models. 

Though our sample is 850\um\, selected, and thus prone to bias towards intrinsically colder systems at $z \sim 2$ \citep{2000eales,2004chapman,2009casey}, our selection of brighter targets at higher $L_{\rm IR}$ is expected to correlate to hotter dust temperatures. \citet{2013lee} use \textit{Herschel}-selected DSFGs, which is unbiased at $z \sim 2$ with respect to dust temperature, and show that there are not many hot dust sources to be found unless they have luminous AGN. While obscured AGN could be the culprit, we cannot discern the presence of AGN without additional data. Therefore, our dataset represents a good sampling of galaxies above characteristic luminosity $L_{\rm IR} > 3 \times 10^{12}$\,\Lsun{} at the $S_{850\rm\,\mu m} > 5.55\,$mJy flux density limit; though there exists a bias with respect to dust temperature for this study, we don't expect many intrinsically hotter sources.

Overall, the measured emissivity spectral index, $\beta$, is consistent (within uncertainties) for the three subsamples at $\beta \sim 2.4$, with a median of $\langle \beta \rangle = 2.4 \pm 0.3$ for the full sample. The median value for the low-$z$ bin is $\langle \beta_{\rm{low-}z} \rangle = 2.41\pm0.13$, for the mid-$z$ bin is $\langle \beta_{\rm{mid}-z} \rangle = 2.26\pm0.16$, and for the high-$z$ bin is $\langle \beta_{\rm{high-}z} \rangle = 2.4\pm0.3$ (see Figure \ref{fig:beta}). Errors on the medians are derived from bootstrapping. The use of ALMA data with single-dish SCUBA-2 data, possibly suffering from confusion boosting, could impact individual derivations of $\beta$ in this sample. Though we have accounted for deboosting as best as possible, the precision to which any individual $\beta$ can be measured can only be improved with matched-beam ALMA data at both frequencies. We further discuss the derived emissivity spectral indices for the sample and caveats of our measurements in Section \textsection4.3.

\begin{figure*}
  \centering
    \includegraphics[width=2\columnwidth]{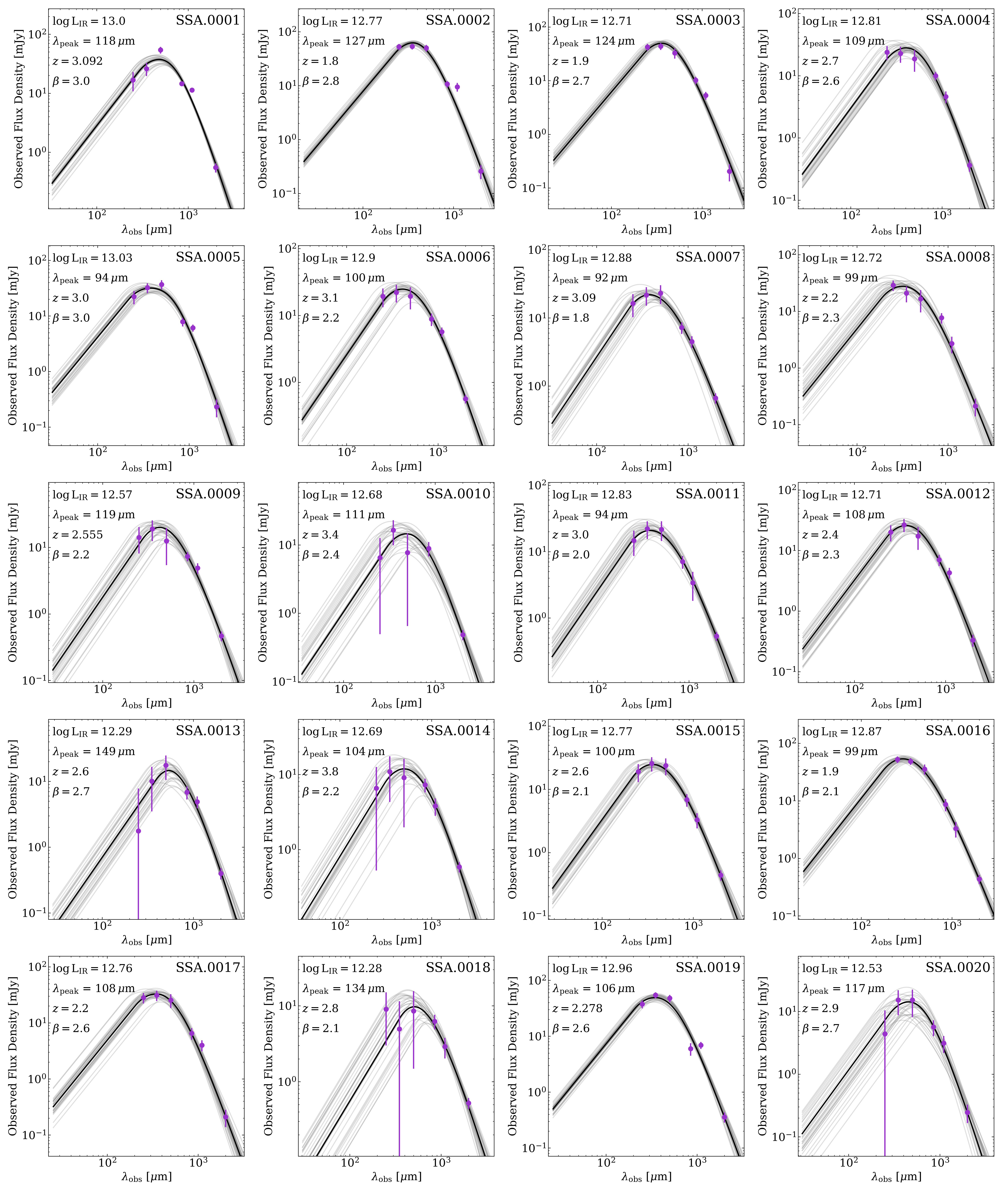}
   \caption{Millimeter photometry (purple points) with the best fit dust SED (black line) for each target in the sample. SEDs are constructed as a modified blackbody with a mid-infrared powerlaw component. For the modified blackbody we assume a general opacity such that $\tau$ = 1 at 200\um, near the intrinsic peak of the dust SED. The uncertainty of the fit is shown via the gray SED fits, drawn randomly from the successful MCMC trials. Measured SED characteristics are noted for each target in their respective panels and given in Table \ref{tab:full}.}
   \label{fig:seds}
\end{figure*}

\begin{figure*}
    \centering
\includegraphics[width=2\columnwidth]{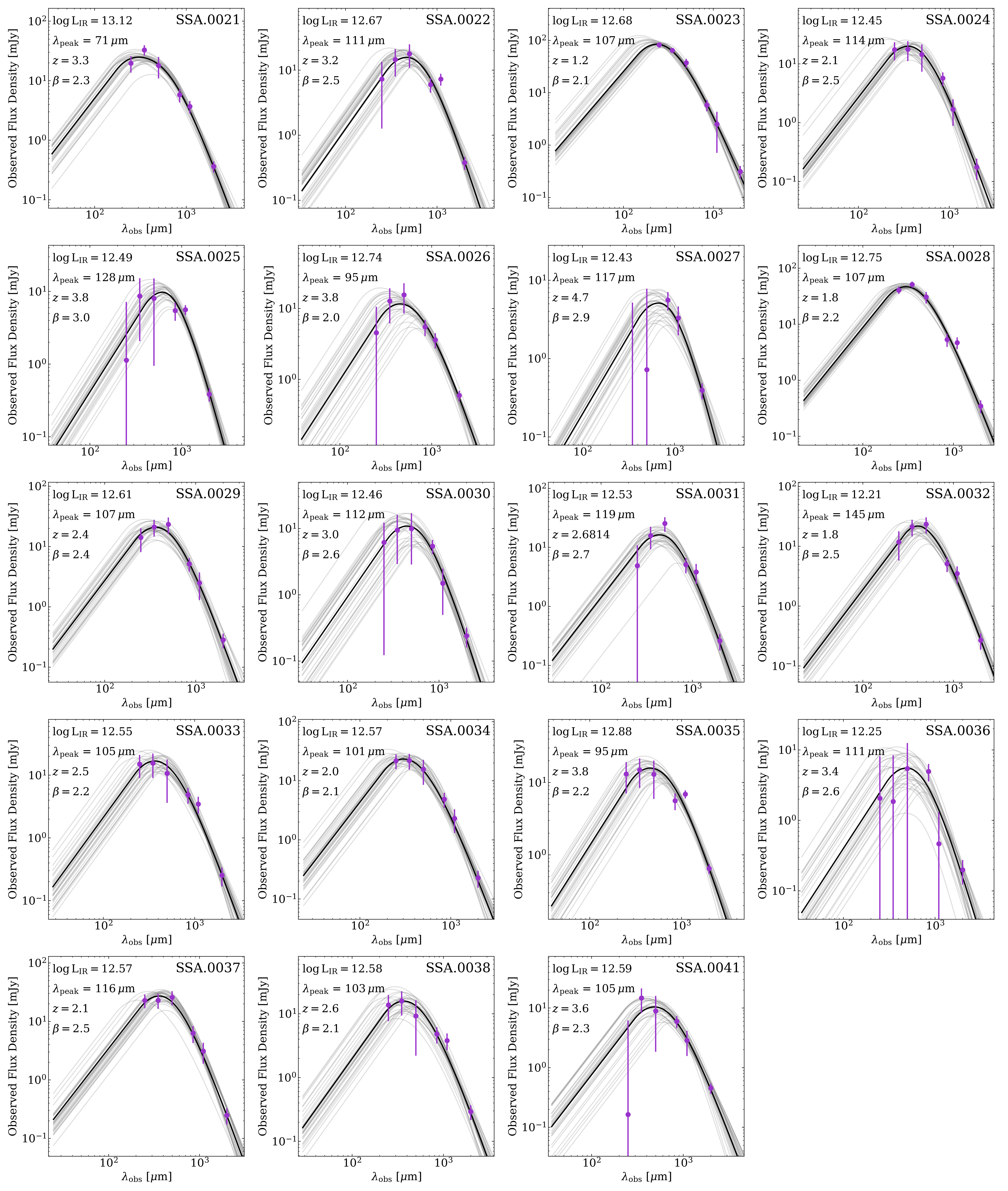}

  --- Continuation of Figure~\ref{fig:seds} ---
\end{figure*}

Using our 2\,mm dust continuum emission, we derive a gas mass for each galaxy by Equation 16 in the Appendix of \citet{2016scoville}. We adopt a single mass-weighted dust temperature of 25\,K \citep[consistent with][]{2016scoville,2019casey} for this calculation rather than our SED measurements of $T_{\rm dust}$, which are luminosity-weighted. As those works argue, a very low fraction of dust mass radiates at hot temperatures relative to the mass of cold dust in a galaxy’s ISM, even when the hot dust contributes substantially to a galaxy’s IR luminosity. Our gas mass estimates show most of our targets are gas-rich with $M_{\rm gas} \sim 10^{11}$\Msun. With gas mass and SFR estimates in hand, gas depletion timescales can be derived in a statistical sense for the sample overall. We find gas depletion times around $\tau \sim 200$ Myr with median and standard deviation $\tau = 220 \pm 160$ Myr, consistent with the majority of $z \gtrsim 1$ DSFGs \citep[e.g.][]{swinbank2014,2020dudz,2021sun}.

The best fit SEDs are presented with photometry overplotted in Figure \ref{fig:seds}, with best fit parameters listed in Table \ref{tab:full}.

\begin{table*}[h!]
\centering
\begin{tabular}{llllllllll} 
\hline
\hline
Name      & $z_{\rm MMpz}$                & $z$ bin  & $z_{\rm OIR}$       & $z_{\rm spec}$     & $\beta$               & log$\, L_{\rm IR}$               & SFR              & $\lambda_{\rm peak}$         & $\rm T_{\rm dust}$*           \\
         &                  &   &     &      &         &    $\rm log_{10}$(\Lsun)         &       \Msunyr     & \um     & K     \\
\hline 
SSA.0001            & $3.3_{-1.6}^{+2.8}$  & Mid-$z$  & $2.85_{-0.98}^{+3.15}$ & $3.092^a$  & $3_{-0.2}^{+0.2}$   & $13.00_{-0.07}^{+0.06}$ & $1500_{-200}^{+200}$ & $118_{-7}^{+7}$      & $41_{-3}^{+3}$   \\
SSA.0002            & $1.8_{-1.4}^{+1.2}$  & Low-$z$  &                        &            & $2.8_{-0.2}^{+0.3}$ & $12.77_{-0.05}^{+0.05}$ & $890_{-90}^{+110}$   & $127_{-8}^{+7}$      & $36_{-3}^{+4}$   \\
SSA.0003            & $1.9_{-1.2}^{+1.1}$  & Low-$z$  &                        &            & $2.7_{-0.2}^{+0.3}$ & $12.71_{-0.06}^{+0.06}$ & $770_{-90}^{+110}$   & $124_{-10}^{+8}$     & $37_{-4}^{+4}$   \\
SSA.0004            & $2.7_{-1.0}^{+1.0}$  & Mid-$z$  &                        &            & $2.6_{-0.3}^{+0.4}$ & $12.81_{-0.10}^{+0.09}$ & $1000_{-200}^{+200}$ & $109_{-13}^{+13}$    & $45_{-6}^{+7}$   \\
SSA.0005            & $2.4_{-1.8}^{+2.5}$  & Mid-$z$  & $3_{-1.67}^{+1.27}$    &            & $3_{-0.3}^{+0.3}$   & $13.03_{-0.08}^{+0.07}$ & $1600_{-300}^{+300}$ & $94_{-9}^{+9}$       & $54_{-5}^{+6}$   \\
SSA.0006            & $3.1_{-0.7}^{+0.7}$  & Mid-$z$  &                        &            & $2.2_{-0.3}^{+0.3}$ & $12.90_{-0.11}^{+0.10}$ & $1200_{-300}^{+300}$ & $100_{-12}^{+12}$    & $49_{-7}^{+7}$   \\
SSA.0007$\dagger$   & $3.2_{-0.5}^{+0.5}$  & Mid-$z$  &                        & $3.09^b$   & $1.8_{-0.2}^{+0.3}$ & $12.88_{-0.11}^{+0.11}$ & $1100_{-300}^{+300}$ & $92_{-9}^{+11}$      & $53_{-6}^{+7}$   \\
SSA.0007.1$\dagger$ &                      &          & $3.7_{-0.19}^{+0.72}$  & $3.0906^b$ &                     &                         &                      &                      &                  \\
SSA.0007.2$\dagger$ &                      &          & $2.90_{-0.32}^{+0.21}$ & $3.0902^b$ &                     &                         &                      &                      &                  \\
SSA.0008            & $1.9_{-0.8}^{+0.9}$  & Low-$z$  & $2.2_{-0.05}^{+0.06}$  &            & $2.3_{-0.4}^{+0.4}$ & $12.72_{-0.12}^{+0.13}$ & $800_{-200}^{+300}$  & $99_{-17}^{+17}$     & $50_{-9}^{+12}$  \\
SSA.0009            & $3.1_{-0.7}^{+0.9}$  & Low-$z$  & $2.2_{-0.32}^{+0.72}$  & $2.555^c$  & $2.2_{-0.3}^{+0.4}$ & $12.57_{-0.16}^{+0.12}$ & $550_{-170}^{+180}$  & $119_{-14}^{+17}$    & $38_{-7}^{+7}$   \\
SSA.0010            & $3.4_{-1.1}^{+2.1}$  & Mid-$z$  &                        &            & $2.4_{-0.5}^{+0.5}$ & $12.68_{-0.20}^{+0.16}$ & $700_{-300}^{+300}$  & $110_{-20}^{+20}$    & $43_{-10}^{+10}$ \\
SSA.0011            & $3_{-0.5}^{+0.4}$    & Mid-$z$  &                        &            & $2_{-0.3}^{+0.3}$   & $12.83_{-0.12}^{+0.11}$ & $1000_{-300}^{+300}$ & $94_{-11}^{+12}$     & $52_{-7}^{+8}$   \\
SSA.0012            & $2.4_{-0.7}^{+0.8}$  & Low-$z$  & $6_{-4.06}^{+0}$**     &            & $2.3_{-0.3}^{+0.4}$ & $12.71_{-0.11}^{+0.10}$ & $800_{-200}^{+200}$  & $108_{-14}^{+13}$    & $44_{-6}^{+8}$   \\
SSA.0013            & $3.5_{-1.2}^{+2.0}$  & Low-$z$  & $2.55_{-0.69}^{+3.45}$ &            & $2.7_{-0.4}^{+0.4}$ & $12.29_{-0.17}^{+0.14}$ & $290_{-100}^{+110}$  & $149_{-17}^{+18}$    & $27_{-5}^{+6}$   \\
SSA.0014            & $3.8_{-0.6}^{+0.6}$  & High-$z$ &                        &            & $2.2_{-0.4}^{+0.5}$ & $12.7_{-0.2}^{+0.2}$    & $700_{-300}^{+400}$  & $100_{-20}^{+30}$    & $46_{-12}^{+11}$ \\
SSA.0015            & $2.6_{-0.5}^{+0.4}$  & Mid-$z$  &                        &            & $2.1_{-0.3}^{+0.4}$ & $12.77_{-0.11}^{+0.10}$ & $900_{-200}^{+200}$  & $100_{-12}^{+13}$    & $49_{-6}^{+7}$   \\
SSA.0016            & $1.9_{-0.4}^{+0.3}$  & Low-$z$  &                        &            & $2.1_{-0.2}^{+0.3}$ & $12.87_{-0.08}^{+0.09}$ & $1100_{-200}^{+300}$ & $99_{-12}^{+11}$     & $49_{-6}^{+8}$   \\
SSA.0017            & $1.9_{-0.8}^{+0.9}$  & Low-$z$  & $2.25_{-0.4}^{+0.2}$   &            & $2.6_{-0.4}^{+0.4}$ & $12.76_{-0.08}^{+0.09}$ & $900_{-200}^{+200}$  & $108_{-14}^{+12}$    & $46_{-6}^{+8}$   \\
SSA.0018            & $3.9_{-0.8}^{+0.9}$  & Mid-$z$  & $2.85_{-0.69}^{+2.7}$  &            & $2.1_{-0.4}^{+0.5}$ & $12.3_{-0.3}^{+0.2}$    & $280_{-130}^{+190}$  & $130_{-20}^{+30}$    & $31_{-9}^{+10}$  \\
SSA.0019            & $2.1_{-1.5}^{+1.5}$  & Low-$z$  & $2.15_{-0.21}^{+0.23}$ & $2.278^d$  & $2.6_{-0.2}^{+0.3}$ & $12.96_{-0.05}^{+0.06}$ & $1350_{-160}^{+190}$ & $106_{-8}^{+8}$      & $47_{-4}^{+5}$   \\
SSA.0020            & $2.9_{-1.1}^{+1.5}$  & Mid-$z$  &                        &            & $2.7_{-0.5}^{+0.5}$ & $12.53_{-0.18}^{+0.16}$ & $500_{-200}^{+200}$  & $117_{-18}^{+16}$    & $41_{-8}^{+9}$   \\
SSA.0021            & $2.4_{-0.6}^{+0.5}$  & Mid-$z$  & $3.3_{-0.21}^{+0.19}$  &            & $2.3_{-0.3}^{+0.4}$ & $13.12_{-0.11}^{+0.10}$ & $1900_{-500}^{+500}$ & $71_{-8}^{+10}$      & $72_{-9}^{+10}$  \\
SSA.0022            & $3.2_{-1.5}^{+3.3}$  & Mid-$z$  &                        &            & $2.5_{-0.4}^{+0.4}$ & $12.67_{-0.18}^{+0.14}$ & $700_{-200}^{+300}$  & $111_{-16}^{+18}$    & $44_{-8}^{+9}$   \\
SSA.0023            & $1.2_{-0.5}^{+0.3}$  & Low-$z$  &                        &            & $2.1_{-0.3}^{+0.3}$ & $12.68_{-0.09}^{+0.14}$ & $700_{-100}^{+300}$  & $107_{-16}^{+13}$    & $44_{-6}^{+9}$   \\
SSA.0024            & $2.1_{-0.9}^{+0.8}$  & Low-$z$  &                        &            & $2.5_{-0.5}^{+0.5}$ & $12.45_{-0.14}^{+0.15}$ & $420_{-120}^{+170}$  & $110_{-20}^{+20}$    & $42_{-8}^{+12}$  \\
SSA.0025            & $3.8_{-1.6}^{+3.5}$  & High-$z$ &                        &            & $3_{-0.5}^{+0.3}$   & $12.5_{-0.2}^{+0.2}$    & $500_{-200}^{+300}$  & $130_{-20}^{+20}$    & $36_{-8}^{+10}$  \\
SSA.0026            & $3.8_{-0.7}^{+0.7}$  & High-$z$ &                        &            & $2_{-0.3}^{+0.4}$   & $12.7_{-0.2}^{+0.2}$    & $800_{-300}^{+400}$  & $95_{-14}^{+20}$     & $51_{-11}^{+10}$ \\
SSA.0027            & $4.7_{-1.7}^{+3.6}$  & High-$z$ &                        &            & $2.9_{-0.6}^{+0.4}$ & $12.4_{-0.2}^{+0.2}$    & $400_{-200}^{+300}$  & $120_{-20}^{+20}$    & $41_{-9}^{+12}$  \\
SSA.0028            & $1.8_{-0.7}^{+0.6}$  & Low-$z$  & $0_{-0}^{+0.01}$**     &            & $2.2_{-0.3}^{+0.3}$ & $12.75_{-0.07}^{+0.08}$ & $840_{-120}^{+160}$  & $107_{-12}^{+10}$    & $45_{-5}^{+7}$   \\
SSA.0029            & $2.4_{-0.7}^{+0.6}$  & Low-$z$  &                        &            & $2.4_{-0.4}^{+0.5}$ & $12.61_{-0.13}^{+0.12}$ & $600_{-200}^{+200}$  & $107_{-17}^{+15}$    & $45_{-7}^{+9}$   \\
SSA.0030            & $3_{-1.1}^{+1.2}$    & Mid-$z$  &                        &            & $2.6_{-0.6}^{+0.6}$ & $12.5_{-0.3}^{+0.2}$    & $400_{-200}^{+200}$  & $110_{-20}^{+20}$    & $43_{-11}^{+12}$ \\
SSA.0031            & $2.8_{-1.5}^{+2.8}$  & Mid-$z$  &                        & $2.6814^e$ & $2.7_{-0.5}^{+0.5}$ & $12.53_{-0.15}^{+0.13}$ & $500_{-150}^{+170}$  & $119_{-16}^{+14}$    & $40_{-6}^{+7}$   \\
SSA.0032            & $2.5_{-0.9}^{+1.0}$  & Low-$z$  & $1.75_{-0.82}^{+4.25}$ &            & $2.5_{-0.5}^{+0.5}$ & $12.21_{-0.12}^{+0.11}$ & $240_{-60}^{+70}$    & $145_{-17}^{+15}$    & $28_{-5}^{+6}$   \\
SSA.0033            & $2.5_{-0.7}^{+0.9}$  & Mid-$z$  &                        &            & $2.2_{-0.4}^{+0.5}$ & $12.55_{-0.18}^{+0.16}$ & $500_{-200}^{+200}$  & $100_{-20}^{+20}$    & $46_{-10}^{+11}$ \\
SSA.0034            & $2.0_{-0.5}^{+0.4}$  & Low-$z$  & $2.05_{-0.2}^{+0.32}$  &            & $2.1_{-0.4}^{+0.5}$ & $12.57_{-0.14}^{+0.15}$ & $600_{-200}^{+200}$  & $101_{-18}^{+18}$    & $48_{-9}^{+12}$  \\
SSA.0035            & $3.8_{-1.5}^{+2.8}$  & High-$z$ &                        &            & $2.2_{-0.4}^{+0.4}$ & $12.9_{-0.2}^{+0.2}$    & $1100_{-500}^{+500}$ & $100_{-10}^{+20}$    & $52_{-12}^{+10}$ \\
SSA.0036            & $3.4_{-1.7}^{+-3.3}$ & High-$z$ &                        &            & $2.6_{-0.7}^{+0.6}$ & $12.3_{-0.4}^{+0.3}$    & $300_{-200}^{+300}$  & $110_{-30}^{+30}$    & $40_{-10}^{+20}$ \\
SSA.0037            & $2.1_{-0.7}^{+0.7}$  & Low-$z$  &                        &            & $2.5_{-0.4}^{+0.5}$ & $12.57_{-0.10}^{+0.11}$ & $550_{-120}^{+150}$  & $116_{-17}^{+16}$    & $40_{-7}^{+9}$   \\
SSA.0038            & $2.6_{-0.8}^{+1.0}$  & Mid-$z$  &                        &            & $2.1_{-0.4}^{+0.5}$ & $12.58_{-0.19}^{+0.16}$ & $600_{-200}^{+300}$  & $103_{-18}^{+20}$    & $47_{-10}^{+12}$ \\
SSA.0041            & $3.6_{-0.8}^{+1.0}$  & Mid-$z$  &                        &            & $2.3_{-0.4}^{+0.5}$ & $12.6_{-0.2}^{+0.2}$    & $600_{-300}^{+300}$  & $110_{-20}^{+20}$    & $46_{-12}^{+11}$ \\
\hline
\hline
\end{tabular}
\caption{Measured and derived characteristics for each galaxy in our sample. Photometric redshift results from \texttt{MMpz} are listed in $z_{\rm MMpz}$, with OIR photometric redshifts quoted as given in \citet{umehata2014} under $z_{\rm OIR}$, and any literature spec-$z$ values listed in $z_{\rm spec}$ with the reference (a: \citet{umehata2014}, b: \citet{2019umehata}, c: \citet{2005chapman}, d: \citet{2012alaghbandzadeh}, e: Cooper et al. in prep). Measured from the best fit SEDs (see Figure \ref{fig:seds}) are emissivity spectral index $\beta$, IR luminosity log$\, L_{\rm IR}$, dust temperature $T_{\rm dust}$, and peak wavelength $\lambda_{\rm peak}$. The star formation rates (SFR) are dervied from $\, L_{\rm IR}$ using the \citet[][]{2012kennicuttevans} scaling. *Note that we assume a general opacity such that $\tau$ = 1 at 200\um, therefore these estimates are only comparable to other works using a similar model. **Reported $z_{\rm OIR}$ from \citet{umehata2014}, which we note is inconsistent with both our FIR results and OIR archival data (see discussion of these sources in \textsection3.1).$\dagger$SSA.0007 is resolved into two components with ALMA (SSA.0007.1 and SSA.0007.2) and each component has both an OIR photo-$z$ and a spec-$z$. However, as the source is unresolved in all other submillimeter bands, we fit the FIR photometry to the sum of the two components (SSA.0007).}
\label{tab:full}
\end{table*}

\section{Discussion}

The goals of this work are to identify the highest-$z$ sources from a sample of luminous DSFGs using 2\,mm observations, characterize 2\,mm emission and FIR SEDs for the sample, provide an independent measurement of the volume density of $z > 3$ DSFGs, and test the practical utility of 2\,mm follow up observations as a useful and efficient redshift filter for large DSFG surveys. Based on FIR SED analysis for our galaxies, we place our sample into context with other DSFG populations, characterize the density of 850\um{}-bright sources in the field, and evaluate the sample's contribution to the CSFRD.

\subsection{Redshift Distribution of $S_{850\rm\,\mu m} > 5.55\,$mJy Sources}

Based on the cumulative probability distribution function (Figure \ref{fig:zbins}), $\sim80\%$ of the aggregate probability density distribution of $S_{850\rm\,\mu m} > 5.55\,$mJy sources is at $z>2$. We categorize 16/39 galaxies as low-$z$, with a minimum redshift solution for an individual source of $z = 1.2^{+0.3}_{-0.5}$ for SSA.0023. We categorize 17/39 sources within the mid-$z$ bin, centered on the protocluster redshift. Based on crossmatching as detailed in Section \textsection 2.3, we find most of our mid-$z$ sample has not been previously cataloged as protocluster members, nor are they expected to be members; rather they have a higher likelihood of being members than the DSFGs in low-$z$ and high-$z$ bins. Indeed, only $\sim5$\% of the volume contained in the mid-$z$ redshift bin corresponds to the protocluster volume itself, leaving substantial room for our DSFGs to lie in the foreground or background population. Taking an overdensity of $\delta_{\rm rare} = 10$ \citep{2016casey} and the protocluster comoving volume within the S2CLS SSA22 coverage (assuming a redshift range informed by \citealt{2018topping}) we may at most expect $\sim 4-7$ of the 17 sources to be protocluster members. Comparing our results to the typical density of $S_{850\rm\,\mu m}>5\,$mJy sources in a general field \citep[][]{swinbank2014,2015umehata}, we find no statistical excess of sources in the mid-$z$ bin; therefore, in the absence of spectroscopic confirmation, we cannot definitively classify these galaxies as protoclusters members. Lastly, we categorize 6/39 sources as high-$z$, with a maximum redshift solution for an individual source of $z = 4.7^{+3.6}_{-1.7}$ for SSA.0027. For the high-$z$ sample, $\sim60\%$ of the aggregate probability density distribution is at $z>4$ (see Figure \ref{fig:zbins}).

While the redshift distribution of galaxies selected at 2\,mm is expected to be relatively high with $\langle z \rangle \approx 3.6$ \citep[][]{2021casey}, the selection of these targets at 850\um{} leads to a lower median redshift, consistent with what has been previously found for 850\um-selected DSFGs \citep[e.g.][]{2020dudz,2021dacunha}. Though we expect the follow up 2\,mm observations to filter out lower redshift sources \citep[][]{2021casey,2021zavala,2022manning}, our source selection here is based on 850\um{}, and thus for this study the redshift distribution is reflective of the 850\um{} DSFG redshift distribution. Previous millimeter studies demonstrate that deeper surveys tend to select for lower redshift DSFGs; in other words brighter DSFGs tend to sit at higher redshifts \citep[e.g.][]{2015bethermin}. Though we focus on only the brightest subset of the population, we find a redshift distribution consistent with other 850\um{}-selected SMG samples, therefore the redshift distribution is not significantly different for this bright subset.

\begin{figure*}[]
    \centering
    \includegraphics[width=2\columnwidth]{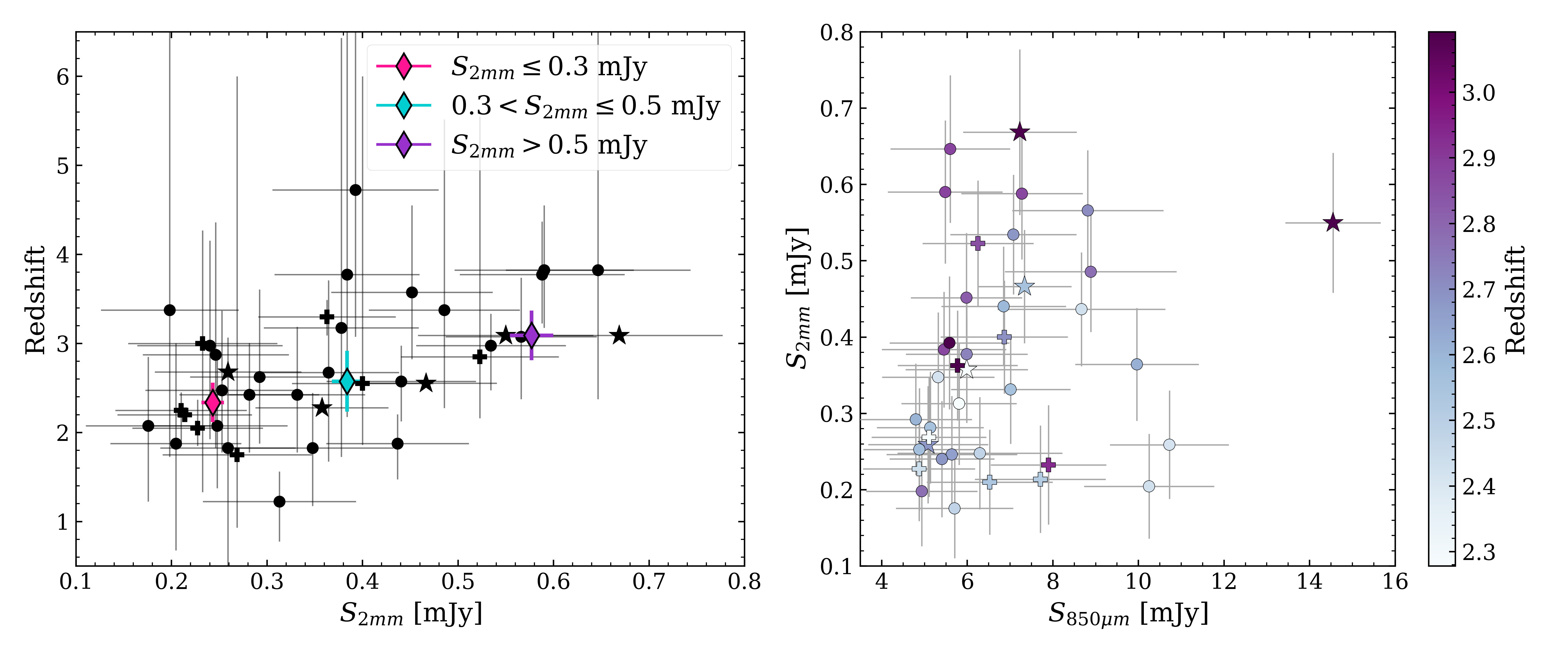}
    \caption{Redshift versus (sub)mm flux density for the $S_{850\rm\,\mu m}>5.55$\,mJy sample. \textit{Left:} Flux density $S_{\rm2\,mm}$ versus redshift showing a positive correlation: sources with higher $S_{\rm2\,mm}$ have higher redshift solutions. The stars indicate sources which have spectroscopic redshifts and pluses denote OIR photometric redshifts, otherwise \texttt{MMpz}-derived redshifts are shown as circles. The average values within given flux density bins are shown as colored diamonds. \textit{Right:} Flux densities $S_{\rm2\,mm}$ versus $S_{850\rm\,\mu m}$ with point color indicating redshift as shown by the colorbar on the right. As before, the stars denote spec-$z$'s, the pluses denote OIR photo-$z$'s, and the circles are mm/FIR photo-$z$'s. Sources with lower flux ratios $S_{850\rm\,\mu m}/S_{\rm2\,mm}$ tend to sit at higher redshifts.}
    \label{fig:z_2mm}
\end{figure*}

We find a positive correlation between 2\,mm flux density and redshift, and sources with lower flux ratios $S_{850\rm\,\mu m}/S_{\rm2\,mm}$ tend to have higher redshifts (see Figure \ref{fig:z_2mm} and Figure \ref{fig:btrax}). We caution that this observation is partially due to the use of 2\,mm flux densities in measuring redshifts for sources lacking other independent redshift constraints; however, it should be noted that removing the 2\,mm flux densities from \texttt{MMpz} redshift determinations still results in a trend between redshift and 2\,mm flux density, though it introduces more scatter. The loose positive correlation between 2\,mm flux density and redshift implies that 2\,mm flux density alone does not constrain redshift, but does suggest sources with higher 2\,mm flux densities tend to sit at higher redshifts. The relation between FIR/mm flux density and redshift becomes more clear with the flux ratio $S_{850\rm\,\mu m}/S_{\rm2\,mm}$; for a source with a given $850\rm\,\mu m$ flux density, a relatively higher 2\,mm flux density tends to result in a higher redshift solution. This trend holds for sources in our sample with $\texttt{MMpz}$-derived or OIR photometric redshifts as well as spectroscopic redshifts.

Our redshift distribution has good agreement with other 850/870\um{}-selected surveys, including \citet[][]{2005chapman}, who conduct an early assessment of the 850\um{}-selected SMG redshift distribution (median $z = 2.2$), and \citet[][]{2011wardlow}, with a similar redshift distribution peaking around $z \sim 2.5$ for their complete, unbiased 870\um{}-selected sample. Similarly, \citet[][]{2017danielson} derive a median redshift of $z = 2.4 \pm 0.1$ for the 52 spectroscopically confirmed SMGs in their sample. \citet[][]{2017danielson} also find the distribution features a high redshift tail, with $\sim$23\% of the SMGs at $z \geq 3$; we find a comparable $\sim33\%$ of the sample at $z \geq 3$. The recent AS2UDS survey \citep{2020dudz} follow-up a comparable sample of $S_{850\rm\,\mu m}>3.6$\,mJy SCUBA-2-selected SMGs in the UKIDSS UDS field with ALMA at 870\um. They find a photometric redshift distribution with median $z = 2.61\pm0.08$ ($1\sigma$ range of $z=1.8 - 3.4$), in good agreement with our sample's median redshift of $z = 2.7 \pm 0.2$ (bootstrapped error on the median) and $1\sigma$ range of $z = 2.1 - 3.4$. In addition, they find $\sim6\%$ of their sample at $z>4$, and only 5/707 ($<<1\%$) sources at $z < 1$. This is broadly consistent with our smaller, brighter sample, as we find zero sources at $z<1$, and find one source ($\sim3 \pm 3\%$ of the sample) formally at $z>4$. Still, we do find a number of galaxies at $z > 3.8$ (5/39 or 13\% of the total sample, and most of the high-$z$ sample). 

We also find consistent results with the ALESS survey \citep{2021dacunha}, a sample of 99 870\um\-selected SMGs followed up with ALMA 2\,mm observations, with a similar flux density limit as AS2UDS of $S_{870\rm\,\mu m} \geq 3.5$\,mJy. To compare to our sample, we take the subset of 25/99 ALESS sources with $S_{870\rm\,\mu m} \geq 5$\,mJy from \citet[][]{2021dacunha}, and find a median redshift and 68\% confidence interval of $z_{\rm med} = 3.2^{+0.4}_{-1.1}$. This is broadly in agreement with our median redshift. Note that the full ALESS sample has a median redshift $z_{\rm med} = 2.8_{-0.8}^{+0.7}$, and includes fainter sources than our sample selection criteria. Still, previous studies note the correlation between $S_{850\rm\,\mu m}$ flux density and redshift is fairly shallow \citep[e.g.][]{2019stach}, so the median redshift for our bright sample is expected to be somewhat consistent with samples pushing to greater depths.

\vfill\null

\subsection{Utility of 2\,mm as a Redshift Filter}

A key motivation in this work is to analyze the utility of 2\,mm in identifying the highest redshift galaxies among a population that lacks spectroscopy or high quality OIR photometric redshifts. This is a core problem in the study of obscured galaxy populations, where the search for redshifts has been the primary bottleneck for the past 20 years \citep*[][]{dsfgcaseyreview}. For example, following up these sources with spectral scans in the millimeter would take over 100 hours with ALMA (at least $\sim 3$ hours per source with overheads), versus the total time spent in this paper, with a total on-source time of just 34.12\,min for 39 sources ($<1$\,min per source). Selection at 2\,mm has been found to effectively select a higher redshift population of DSFGs, as shown by simulations in \citet{2014zavala} and by the blind 2\,mm MORA survey in \citet{2021casey}, with $\langle z\rangle = 3.6^{+0.4}_{-0.3}$ for their sample. Note that while similar long wavelength observations (for example at 3\,mm) are also informative, we opt for 2\,mm observations since sources are $\sim4-5\times$ brighter at 2\,mm and therefore 2\,mm follow-up is more observationally efficient. 

Our hypothesis is that 2\,mm imaging can provide an efficient and targeted method for identifying high-$z$ candidates for follow-up as a zeroth order redshift sorting of 850\um{}-bright sources. In this study, we are limited to sources that are detected at 850\um{} by construction, given our selection criteria. Besides ALMA, other instruments with 2\,mm imaging capabilities -- such as the NOrthern Extended Millimeter Array (NOEMA) and the upcoming TolTEC instrument on the Large Millimeter Telescope (LMT) -- can be used for targeted 2\,mm follow-up as well. Notably, TolTEC will offer high-angular-resolution simultaneous observations at 1.1, 1.4, and 2\,mm \citep[][]{toltec}, allowing redshift-sorting color cuts to be made directly without ALMA follow-up.

Definitively assessing the utility of 2\,mm imaging for redshift filtering requires spectroscopic redshifts, but in the absence of spectroscopic data we can evaluate the impact of 2\,mm data on our derivation of redshifts relative to the other bands for which we have data. To derive a photo-$z$, \texttt{MMpz} uses all flux densities for which galaxies have measurements; is our redshift solution primarily driven by 2\,mm flux density, which would render the correlation of 2\,mm flux density and redshift in Figure \ref{fig:z_2mm} unsurprising?

To evaluate the set of photometry that has the most influence for \texttt{MMpz} redshift solutions, we generate simulated galaxy FIR/mm photometry. We specifically test the sensitivity of \texttt{MMpz} on the addition or absence of different bands by generating mock SEDs using the same methodology as \citet[][]{mmpz} where sources of different fixed redshifts are assigned realistic flux densities with associated uncertainties matched to those of this dataset. We then fit redshifts with \texttt{MMpz} for subsets of the data including the full data set: (a) SPIRE 250\um{}/350\um{}/500\um{}, SCUBA-2 850\um{}, AzTEC 1.1\,mm, ALMA 2\,mm; (b) just SCUBA-2 850\um{} and ALMA 2\,mm; and just (c) SPIRE 250\um{}/350\um{}/500\um{} and SCUBA-2 850\um{}. \citet[][]{mmpz} notes, and we find using the same process described therein, that using only 850\um{} and 2\,mm photometry as in (b) -- two points that should exclusively probe the Rayleigh-Jeans tail for most redshifts and temperature -- is insufficient. This is because with only these two points, the peak of the dust SED is not bracketed, and a redshift cannot be well constrained. SPIRE 250\um{}/350\um{}/500\um{} data is needed to constrain the dust peak for $z \lesssim 6$, where even non-detections provide very useful (though loose) constraints on the SED peak. 

From these simulated SEDs, the addition of the 2\,mm point -- provided $\sim500$\um\, constraints exist -- improves accuracy from $\Delta z/(1+z) = 0.3$ (without the 2\,mm data) to $\Delta z/(1+z) = 0.2$ (with the 2\,mm data), in particular for galaxies at $z \gtrsim 3.5$. This demonstrates that given \textit{Herschel} data, the 2\,mm data is the next most impactful data for improved accuracy of the redshift solution, especially for high redshift sources. For example, the probability of a hypothetical source with $z_{\rm MMpz} = 4.1$ lying at $z>3.5$ increases from 81\% without the 2\,mm constraint, to 96\% with the 2\,mm constraint. Note that the lower $\Delta z/(1+z)$ measured for the sample compared to the predicted $\Delta z/(1+z)$ from the mock SED analysis is likely because there is only a small subsample of sources with spectroscopic redshift information. With a larger sample of spectroscopically confirmed sources we'd likely see a wider range of SEDs represented, and estimate a sample $\Delta z/(1+z)$ more consistent with the prediction. 

While 2\,mm data does improve the accuracy of redshift constraints, they do not significantly change the precision of those redshift solutions. Quantitatively, the average breadth of redshift PDFs for our sample is $\delta z = 1.1 - 1.2$ with or without the 2\,mm data folded into the photometry. This is not unexpected, as the breadth of the redshift PDF generated from millimeter data is dominated by the intrinsic spread in the $L_{\rm IR}$-$\lambda_{\rm peak}$ relation (discussed at greater depth in \citealt{mmpz}).

\subsection{Emissivity Spectral Index}

With 2\,mm flux densities in hand for a large sample of bright 850\um{}-selected DSFGs, we have a unique dataset with which we analyze the galaxy-integrated slope of the Rayleigh-Jeans tail of blackbody emission. The slope is governed by the physical quantity $\beta$, the emissivity spectral index, or the frequency dependence of dust grain emissivity per unit mass. Given that measurement of $\beta$ has very weak redshift dependence, this quantity can be measured even without reliable redshift information \citep{mmpz}. Still, drawing a physical interpretation for a measured $\beta$ is not possible for spatially unresolved high redshift galaxies due to the complexity and heterogenity of the ISM. This is especially true for sources lacking secure redshifts and with somewhat limited FIR/mm photometry.

Nevertheless, our data can be used to compare galaxy-integrated Rayleigh-Jeans slopes to other high-$z$ samples and commonly adopted literature values of $\beta$ for high-$z$ datasets of similar quality. Here, our finding of $\langle \beta \rangle = 2.4\pm0.3$ suggests that the distribution of integrated $\beta$ indices skews high, in line with other recent works, including a well-studied galaxy in SSA22 with $\beta = 2.3$ \citep[][]{2018kato} as well as other samples with dust continuum data at $\lambda_{\rm obs} \gtrsim 2\,$mm \citep[][]{2019jin,2021casey}. The aggregate best fit SED results for our sample (see Table \ref{tab:full}) shows relatively steep $\beta$ values ranging from $1.8 < \beta < 3.0$, all steeper than the standard value often adopted in the literature, typically $\beta = 1.8$ \citep[e.g.][]{2016scoville}, justified by measurements of the $\beta$ from Milky Way’s ISM \citep[e.g.][]{2009paradis,2011planck}. Nevertheless, while our median $\beta$ skews high at $\langle \beta \rangle = 2.4\pm0.3$, it remains consistent with theoretical predictions for interstellar dust models \citep[e.g.][]{1984drainelee,2015kohler}, which predict $1 < \beta < 2.5$ depending on grain composition. 

A steeper $\beta$ could result from a fundamental difference in grain composition or size \citep[][]{2001chihara,2013mutschke,2020inoue}, although testing this would require probing the ISM with high spatial resolution and sensitivity. Many of the caveats influencing observed $\beta$ measurements -- including geometric effects, variations in optical depth, and temperature distributions -- would flatten rather than steepen the SED at long wavelengths.

\begin{figure}[]
    \centering
    \includegraphics[angle=0,trim=0in 0in 0in 0in, clip, width=0.5\textwidth]{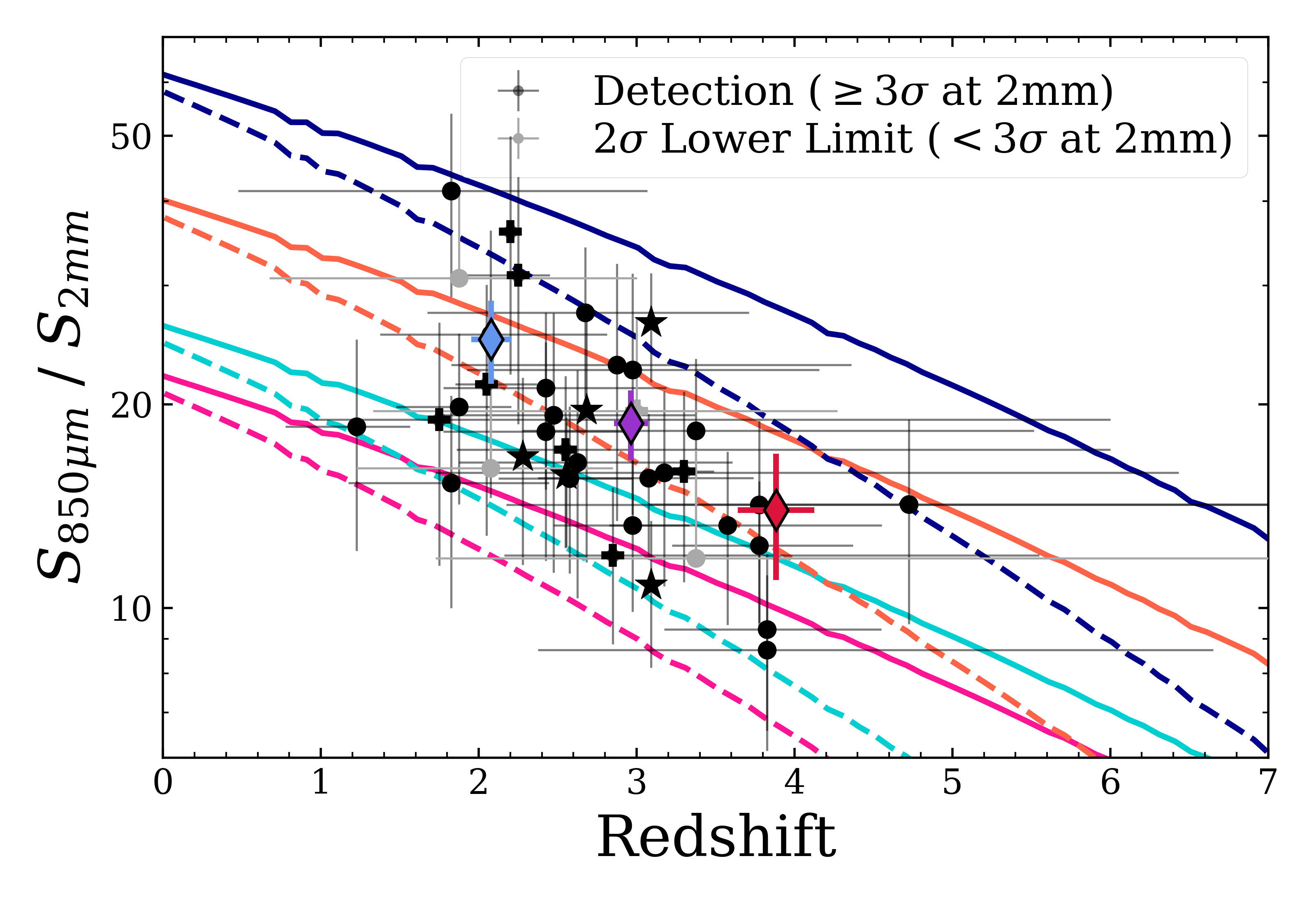}
    \caption{Flux density ratio $S_{850\rm\,\mu m} / S_{\rm 2\,mm}$ versus redshift for the full sample; while 850\um{} flux densities are SCUBA-2 based, they have been deboosted for confusion noise and the errors here reflect that additional source of uncertainty. Redshifts are plotted with spec-$z$'s as stars, OIR photo-$z$'s as pluses, and mm/FIR photo-$z$'s as circles. Detected targets at 2\,mm are denoted as secure detections ($\geq3 \sigma$ at 2\,mm, black points) or plotted as $2\sigma$ lower limits for non-detections (detected at $<3 \sigma$ at 2\,mm, in gray). Averages for each redshift bin (low-$z$ in blue, mid-$z$ in purple, high-$z$ in red) with bootstrapped errors are shown as diamonds. Overlaid as colored lines are SED tracks, each a modified blackbody with mid-infrared power law for a given combination of emissivity spectral index ($\beta$ = 1.8, 2.0, 2.5, and 3.0 in pink, cyan, orange, and navy, respectively) and dust temperature (given here as the observable, $\lambda_{\rm peak}$), with solid lines for $\lambda_{\rm peak} = 90$\um$\,$ and dashed for $\lambda_{\rm peak} = 120$\um. Our sample is largely consistent with steeper slopes consistent with SED tracks for $\beta \approx 2.0 - 2.5$, with some variation (the median $\beta$ measured is 2.4).}
    \label{fig:btrax}
\end{figure}

Figure \ref{fig:btrax} indicates the millimeter color distribution for the sample, represented by observed flux density ratio $S_{850\rm\,\mu m} / S_{\rm 2\,mm}$.  We compare the millimeter colors to SED tracks constructed from a modified blackbody spanning $1.8 < \beta < 3.0$ and 90\um $< \lambda_{\rm peak} < 120$\um. The sources follow a general trend of $S_{850\rm\,\mu m} / S_{\rm 2\,mm}$ with redshift aligned with the shape of the model SEDs. Though this trend may be due to the use of millimeter photometry to derive redshifts, we note that \texttt{MMpz} fixes $\beta = 1.8$ and that modifications of $\beta$ do not significantly impact the output redshift probability density distributions. The majority of sources have flux density ratios suggestive of high $\beta > 1.8$, with the sample largely consistent with SED tracks for $\beta \approx 2.0 - 2.5$ (cyan and orange tracks in Figure \ref{fig:btrax}). This aggregate high $\beta$ result suggests that steeper Rayleigh-Jeans slopes, i.e. higher $\beta$, should likely be applied for similar high-$z$ DSFGs in the literature in future works.

To test the impact of SED fitting approach on our results, we also fit our SEDs using isothermal modified blackbody models as in \citet[][]{2021dacunha}. Given the known degeneracies and bias from opacity assumptions, we toggled $\lambda_0$ over multiple fits. For $\lambda_0=200$\um{} (same assumption as in our modelling), we find a median of $\beta=2.1\pm0.5$, statistically consistent within error with our $\beta \approx 2.4$ measurement. Note that while in both SED fitting routines we assume isothermal dust, a primary difference between methods is the Drew \& Casey (submitted) model accounts for the range in temperature on Wien’s portion of the SED with a power-law, whereas \citet[][]{2021dacunha} do not attempt to fit any points below 70\um{} rest-frame with their simple single-component model. The discrepancy in results from these two methods demonstrates that care should be taken when directly comparing emissivity spectral indices obtained using different fitting methods. To this point, using identical fitting methods we do find a higher median emissivity spectral index for our sample ($\beta=2.1\pm0.5$) than the ALESS sample ($\beta = 1.9 \pm 0.4$) \citet[][]{2021dacunha}. 

Though we have accounted for confusion noise as best as possible by using deboosted flux densities from SCUBA-2, measuring a higher $\beta$ due to combining single dish 850\um\,data with interferometric ALMA 2\,mm data is a concern. Still, \citet[][]{2021dacunha} use multiwavelength ALMA data and demonstrate $\beta \approx 2$ for high-$z$ DSFGs, with a median $\beta = 1.9 \pm 0.4$ derived for their sample. It is possible the bias from single dish data could be responsible for the difference between a $\beta \approx 2$ result and our higher $\beta \approx 2.4$ measurement. If it is, it would also mean that single-dish deboosted flux densities are not quite `deboosted' enough. Large samples with redshifts and all ALMA data on the Rayleigh-Jeans tail are needed to take better measurements with respect to confusion noise and multiplicity in detected sources.

\subsection{CSFRD Contribution}

\begin{figure}[t]
    \centering
    \includegraphics[angle=0,trim=0in 0in 0in 0in, clip, width=0.5\textwidth]{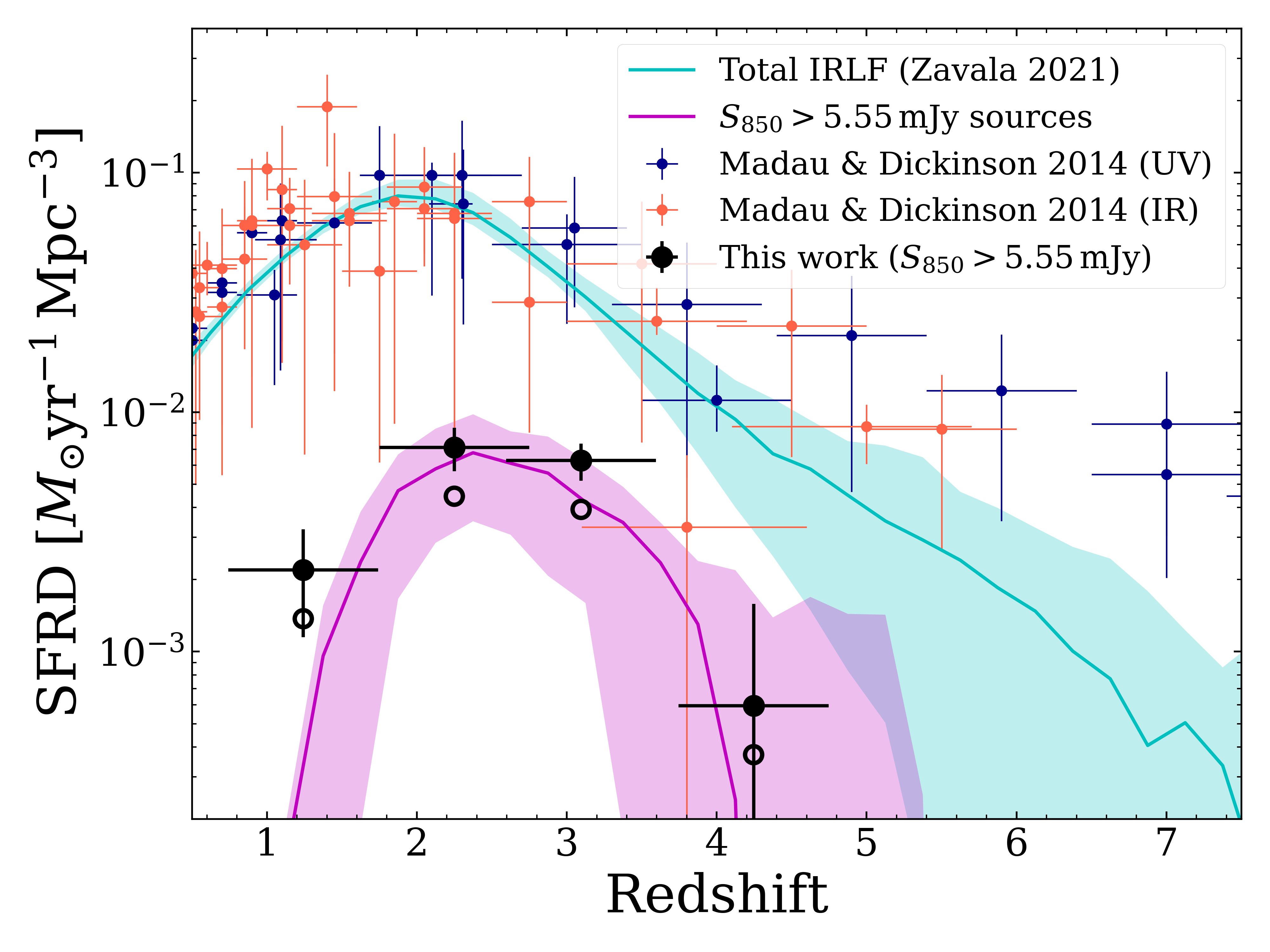}
    \caption{CSFRD as measured at rest-frame UV (blue) and IR-mm (orange) from \citet[][]{2014madaudickinson}. While rest-frame UV measurements now reach $z>10$ thanks to deep HST NIR imaging campaigns, in contrast, surveys of obscured emission in galaxies remain uncertain at $z > 5$. The black points show the data, which lacks precision to constrain the CSFRD at high-$z$. The cyan region shows the IRLF from \citet[][]{2021zavala}, and the magenta region is derived from this IRLF but only counts sources with $S_{850\rm\,\mu m}>5.55\,$mJy, corresponding to the $S_{850\rm\,\mu m}>5\,$mJy and $>5\sigma$ selection for our sample at the SSA22 S2CLS depth. The filled black points show the CSFRD estimates from our SSA22 data corrected for incompleteness, while the open circles are the same data without correcting for incompleteness.}
    \label{fig:sfrd}
\end{figure}

We determine the sample's estimated contribution to the cosmic star formation rate density (CSFRD), as shown in Figure \ref{fig:sfrd}, to compare with other similar estimates of DSFG samples in the literature. While we do not expect the presence of the protocluster to impact our measurements, we verify this by analyzing the relative density of our sources compared to field samples. CSFRD estimates from this sample are measured using a 100-trial Monte Carlo (MC) to sample the redshift and SFR probability density distributions for each of the 39 SSA22 sources. We calculate a comoving volume over the survey area (S2CLS coverage of SSA22 spans 0.28 deg$^2$) and redshift range based on discrete redshift bins (from $1 < z < 5$ with width $\Delta z = 0.5$), then divide the total SFR of all DSFGs in our sample by that volume when their draws fall in the given $z$-bin. The total CSFRD is measured for each MC trial, and from these measurements we take the mean and standard deviation for all trials to find the estimate for the CSFRD contribution from $1 < z < 5$ for this $S_{850\rm\,\mu m} > 5.55$\,mJy sample.

Given the overall rarity of $S_{850\rm\,\mu m} > 5.55$\,mJy sources among DSFGs, we estimate the impact of cosmic variance, and of incompleteness due to shallower S2CLS data of SSA22 relative to other fields like the UDS \citep[][]{s2cls} and COSMOS \citep[][]{2019simpson} SCUBA-2 maps. We conduct MC trials to count the number of sources in randomly placed SSA22-sized areas (0.28 deg$^2$) over UDS (0.74 deg$^2$) and COSMOS (1.94 deg$^2$); the 1$\sigma$ 850\um{} depths are 0.9\,mJy for UDS, 1.2\,mJy for COSMOS, and 1.2\,mJy for SSA22. We first enforce the same criteria used to select our SSA22 sources for the COSMOS/UDS sources: $S_{850\rm\,\mu m}>5\,$mJy and $>5 \sigma$. A first round of MC trials is run directly from the COSMOS/UDS catalogs for sources that meet these criteria. In this round, we find an average of $62^{+24}_{-30}$ sources/SSA22-sized field. This exceeds the 39 observed sources in SSA22 by a factor of $1.6^{+0.6}_{-0.8}$; though suggestive of an underdensity of bright DSFGs in SSA22, this does not account for the higher rms noise in SSA22 relative to UDS/COSMOS. Thus this factor is more representative of incompleteness. For the next round of trials, we artificially boost the rms of the UDS/COSMOS maps using the rms distribution of the SSA22 map to simulate the shallower SSA22 map. We do this by drawing rms values from the SSA22 map at random positions. After this elevation of noise for the COSMOS/UDS samples, we find an expectation value of $44^{+13}_{-14}$ sources/SSA22-sized field fulfilling our bright DSFG criteria. This is in line with our observed number of 39 DSFGs.

From this sampling of UDS/COSMOS maps, we determine that our SSA22 sample is not cosmologically over- or under-dense, but it is incomplete (in its representation of all $S_{850\rm\,\mu m}>5.55\,$mJy sources) by a factor of $1.6^{+0.6}_{-0.8}$. While this is consistent with being complete, it does reveal a systematic offset that impacts our results based on the depth of SSA22 SCUBA-2 coverage, thus we adjust our CSFRD measurement accordingly by multiplying by the derived incompleteness factor.

We also run similar MC trials on simulated data to compare to our observational results. With the semiempirical model from \citet[][]{2020popping}, we count the number of $S_{850\rm\,\mu m}>5\,$mJy sources in randomly placed SSA22-sized areas (0.28 deg$^2$), and find $61^{+8}_{-15}$ sources/SSA22-sized field. From similar MC trials on SHARK light cone results \citep[][]{2020lagos} we find $48^{+12}_{-9}$ sources/SSA22-sized field. These results are consistent with the observational results from UDS \& COSMOS S2CLS, supporting our estimates of incompleteness and cosmic variance for our sample from the real data. 

The resulting CSFRD estimate for our sample is shown in Figure \ref{fig:sfrd}. This estimate is compared to the expected contribution by obscured galaxies to the CSFRD drawn from the IR luminosity function (IRLF) constraints in \citet{2021zavala}. We compare directly to the expected CSFRD for $S_{850\rm\,\mu m}>5.55\,$mJy from their work, corresponding to the $S_{850\rm\,\mu m}>5\,$mJy and $>5\sigma$ selection for our sample at the $1\sigma$ depth ($\sim1.2\,$mJy) of the S2CLS data for SSA22. As shown in Figure \ref{fig:sfrd}, our incompleteness-corrected results are consistent with the \citet[][]{2021zavala} IRLF model within uncertainties.

Although we enforce a higher flux density cutoff than their sample, CSFRD estimates for our sample are comparable to those for the AS2UDS sample \citep[][]{2020dudz}. For our sample, we find the contribution to the CSFRD peaks around $z \sim 2.4$, slightly higher redshift than the total CSFRD peak at $z \sim 2$ from \citet[][]{2014madaudickinson}. Our 850\um{}-bright, obscured sources contribute $\sim10$\% (ranging 8-13\%) to the cosmic-averaged CSFRD from $2<z<5$.

Using a similar MC trials technique, we also compute volume density. For a 100-trail MC, we find the volume density of DSFGs from $4 < z < 5$ to be $6.3\times10^{-7}\rm\,Mpc^{-3}$ based on our sample. Note that this volume density is very specifically linked to our flux density cutoff, as different cuts will result in more or fewer sources (Long et al. in prep).

\section{Summary}

In this paper we present ALMA 2\,mm imaging for a complete sample of 39 SCUBA-2 detected DSFGs in SSA22 selected for $S_{850\rm\,\mu m}>5\,$mJy at $>5\sigma$ in S2CLS, corresponding to a $S_{850\rm\,\mu m}>5.55\,$mJy limit given the survey depth. We detect 35/39 sources at $S_{2\rm{mm}} >3\sigma$, where our sensitivity is rms $\sim$ 0.08 mJy beam$^{-1}$ on average. With multiwavelength (sub)millimeter data for the sample from \textit{Herschel}/SPIRE, SCUBA-2, AzTEC, and ALMA, we characterize IR SEDs and measure and derive properties including IR luminosity, star formation rate, and emissivity spectral index. For each galaxy, we also estimate a millimeter photometric redshift with \texttt{MMpz}. Our main results are as follows:

\begin{itemize}
    \item Based on our photometric redshifts and literature spectroscopic redshifts, we find a redshift distribution consistent with other 850\um{}-selected SMG samples, with median $\langle z \rangle = 2.7 \pm 0.2$ despite these sources being representative of only the brightest subset of the 850\um{}-selected population (often hypothesized to sit at higher redshifts). We also categorize 6/39 sources as high-$z$ (all with redshift solutions $z \gtrsim 3.4$), to be spectroscopically confirmed in future follow-up observations.
    
    \item Provided the existence of $250 - 500$\um{} photometry (i.e. from SPIRE) that brackets the dust SED peak for the majority of DSFGs, we find that the 2\,mm photometric constraints are the next most impactful for refining redshift solutions for these sources. With available redshift constraints, we generally find a positive correlation between redshift and 2\,mm flux density. This 2\,mm data point is especially useful for high redshift sources at $z>3.5$, where the addition of the 2\,mm point improves accuracy from $\Delta z/(1+z) = 0.3$ (without the 2\,mm data) to $\Delta z/(1+z) = 0.2$ (with the 2\,mm data). 
    
    \item Our sample has broadly steep emissivity spectral indicies with median $\beta = 2.4 \pm 0.3$. Still, while the aggregate $\beta$ skews high, measurements for any individual $\beta$ would be improved with matched beam ALMA data at both frequencies, with respect to confusion noise and multiplicity in detected sources.

    \item For our sample of $S_{850\rm\,\mu m}>5.55\,$mJy sources, we estimate the contribution to the cosmic-averaged CSFRD is $\sim10$\% (ranging 8-13\%). We find that this sample of DSFGs in SSA22 is representative of the cosmic average density, within uncertainties, and not impacted by the known protocluster structure in the field.
    
\end{itemize}

Our study employs ALMA 2\,mm imaging to filter out lower redshift DSFGs, advancing efforts to take a complete census of DSFGs at early epochs. While we do find that sources with higher 2\,mm flux densities tend to sit at higher redshifts, spectroscopic follow up is needed to verify the efficacy of the 2\,mm redshift filtering technique. Future 2\,mm coverage of large samples of DSFGs with ALMA, TolTEC, NIKA-2, and other 2\,mm imaging facilities will be a useful tool to improve the efficiency of identifying the highest redshift obscured galaxies.

\section*{Acknowledgements}

The authors would like to extend their gratitude to the anonymous referee for their constructive comments, which helped improve the manuscript. The authors also thank Mark Swinbank for providing SPIRE imaging, Hideki Umehata for sharing AzTEC 1.1\,mm imaging and catalog data, and Claudia Lagos for providing SHARK model lightcone results. ORC thanks the UT Austin Astronomy Department and the Cox Board of Trustees for support through the Dean's Excellence Fellowship. CMC thanks the National Science Foundation for support through grants AST-1814034 and AST-2009577, and the Research Corporation for Science Advancement for a 2019 Cottrell Scholar Award, sponsored by IF/THEN, an initiative of Lyda Hill Philanthropies. E.d.C. gratefully acknowledges the Australian Research Council as the recipient of a Future Fellowship (project FT150100079) and the ARC Centre of Excellence for All Sky Astrophysics in 3 Dimensions (ASTRO 3D; project CE170100013).

This paper makes use of the following ALMA data: ADS/JAO.ALMA\#2019.1.00313.S. ALMA is a partnership of ESO (representing its member states), NSF (USA) and NINS (Japan), together with NRC (Canada), MOST and ASIAA (Taiwan), and KASI (Republic of Korea), in cooperation with the Republic of Chile. The Joint ALMA Observatory is operated by ESO, AUI/NRAO and NAOJ. The National Radio Astronomy Observatory is a facility of the National Science Foundation operated under cooperative agreement by Associated Universities, Inc.

This research made use of the following software: spyder \citep{spyder}, astropy \citep{astropy}, Matplotlib \citep[][]{matplotlib}, NumPy \citep{numpy}, SymPy \citep{sympy}, SciPy \citep[][]{scipy}, pandas \citep{pandas-proceedings,pandas-software}.

\bibliographystyle{aasjournal}
\bibliography{MyPubs}

\end{document}